\documentclass{article}

\usepackage{amssymb}
\usepackage{amsthm}
\usepackage{fancyhdr}
\usepackage{authblk}
\usepackage[title]{appendix}
\usepackage{array}
\usepackage[tableposition=top]{caption}
\usepackage{dsfont}
\usepackage{graphicx}
\usepackage{mathtools}
\usepackage{multicol}
\usepackage{multirow}
\usepackage{pgfplots}
\usepackage{tikz}
\usepackage{xcolor}
\usepackage{enumitem}
\setlist[itemize]{noitemsep}
\setlist[enumerate]{noitemsep}
\usepackage{arxiv}
\usepackage{booktabs}       
\usepackage{nicefrac} 

\usepackage{enumitem}
\usepackage{algorithm}
\usepackage{algpseudocode}
\usepackage{xpatch}

\usepackage[superscript,biblabel]{cite}

\usepackage[OT1]{fontenc}

\usepackage{url}
\usepackage{hyperref}

\usepackage[capitalise]{cleveref}

\usepackage{amsmath}
\usepackage{arxiv}

\pgfplotsset{compat=1.4}

\pgfplotsset{
    jitter/.style={
        x filter/.code={\pgfmathparse{\pgfmathresult+(rnd-.5)*#1}}
    },
    jitter/.default=0.5
}

\usetikzlibrary{hobby, graphs, decorations.markings, arrows, shapes, positioning, mindmap, backgrounds, calc, matrix, patterns, bayesnet, patterns.meta}

\usepgfplotslibrary{statistics, fillbetween, external}

\makeatletter
\xpatchcmd{\algorithmic}{\itemsep\z@}{\itemsep=0.8ex plus2pt}{}{}
\makeatother

\makeatletter
\renewcommand{\fnum@algorithm}{\fname@algorithm}
\makeatother
\algrenewcommand\algorithmicrequire{\textbf{Input:}}
\algrenewcommand\algorithmicensure{\textbf{Output:}}
\algnewcommand{\LeftComment}[1]{\Statex \(\triangleright\) #1}

\usepackage{pgfplots}  
\usepackage{pgfplotstable}  
\pgfplotsset{compat=1.17}  

\usetikzlibrary{hobby, graphs, decorations.markings, arrows, shapes, positioning, mindmap, backgrounds, calc, matrix, patterns, bayesnet, patterns.meta}

\usepgfplotslibrary{statistics, fillbetween, external}

\newtheorem*{theorem*}{Theorem}

\newtheorem*{corollary*}{Corollary}

\newtheorem*{rmk}{Remark}

\newcommand{\ind}{\mathds 1}

\newcommand{\nA}{N_{ \mathcal A }}

\newcommand{\nB}{N_{ \mathcal B }}

\newcommand{\nS}{N_{ \text{synth} }}

\newcommand{\hatFDP}{\widehat{\text{FDP}}}
\newcommand{\hatFDPS}{\widehat{\text{FDP}}_{\text{synth}}}

\begin{document}

\title{ False Discovery estimation in Record Linkage }

\renewcommand\Authsep{, }
\renewcommand\Authands{, and }

\author{Kayan\'{e} Robach\thanks{Corresponding Author: \texttt{k.c.robach@amsterdamumc.nl}}}

\author{Michel H.\ Hof}

\author{Mark A.\ van de Wiel}

\affil{Department of Epidemiology and Data Science, \protect\\ Amsterdam UMC location Vrije Universiteit Amsterdam, De Boelelaan 1117, 1081 HV Amsterdam, The Netherlands}
\affil{Amsterdam Public Health, Methodology, The Netherlands}

\maketitle

\abstract{Integrating data from multiple sources expands research opportunities at low cost. However, due to different data collection processes and privacy constraints, unique identifiers are unavailable. Record Linkage (RL) algorithms address this by probabilistically linking records based on partially identifying variables. Since these variables lack the strength to perfectly combine information, RL procedures yield an imperfect set of linked records. Therefore, assessing the false discovery proportion (FDP) in RL is crucial for ensuring the reliability of subsequent analyses. In this paper, we introduce a novel method for estimating the FDP in RL for two overlapping data sets. We synthesise data from their estimated empirical distribution and use it along with real data in the linkage process. Since synthetic records cannot form links with real entities, they provide a means to estimate the amount of falsely linked pairs. Notably, this method applies to all RL techniques and across diverse settings where links and non-links have similar distributions---typical in complex tasks with poorly discriminative linking variables and multiple records sharing similar information while representing different entities. By identifying the FDP in RL and selecting suitable model parameters, our approach enables to assess and improve the reliability of linked data. We evaluate its performance using established RL algorithms and benchmark data applications before deploying it to link siblings from the Netherlands Perinatal Registry, where the reliability of previous RL applications has never been confirmed. Through this application, we highlight the importance of accounting for linkage errors when studying mother-child dynamics in healthcare records.}


\section{Introduction}\label{sec 1 introduction}

The development of record linkage (RL) methodologies stems from the concept of assembling a `book of life' for individuals to improve public health and human wellness\cite{bookoflife}. This approach enhances research by integrating data from multiple sources to address novel questions. Today, privacy regulations drive the advancement of RL methods to accurately combine observations collected on different occasions using partially identifying variables (e.g.\ birth year, postal code). These variables are prone to errors and possess limited discriminative power due to a restricted number of unique values, which pose challenges in the linkage process, requiring robust error measurement tools.

Linked data contain two types of errors: falsely linked pair and missed link. The implications of linkage errors depend on the linkage structure\cite{datalinkagequal2020, reflexRL_2020}. We focus on settings where linkage is used to construct the study population which corresponds to the intersection between two sets of records. In that context, the falsely linked pairs introduce noise, diluting true signals thus potentially biasing analyses. They lead to an overestimation of the sample size, creating a misleading impression of the amount of information available in the data. In contrast, missed links reduce the sample size, increase variance and diminish statistical efficiency. The false negative proportion (FNP) quantifies missed links, which are particularly difficult to detect due to registration errors (e.g.\ missing values, typographical errors) or real-world changes over time (e.g.\ individuals relocating). Estimating the FNP requires accounting for such discrepancies, as records referring to the same individual may contain variations. Addressing falsely linked pairs through the false discovery proportion (FDP) is more feasible. 

Probabilistic RL methods facilitate FDP estimation, either by leveraging agreement probabilities in linkage variables under the Fellegi-Sunter framework\cite{belin_rubin_1995, FS_RL_1969, measurementError_2019} or by using posterior linkage probabilities\cite{fastLink_paper_2019, sadinle_bay_bipartite_RL_2017, flexrlpackage_2024}. A cutoff rule classifies pairs as linked or non-linked, from which an FDP estimate can be derived\cite{fastLink_paper_2019, fdr_genomewide_2003}. Nonetheless, evaluating the FDP remains challenging because such an estimate is usually not available in the open-source implementations of RL, and its reliability is contingent on the RL model used. Hence it is sensitive to the low discriminative power of linkage variables, registration errors, intricate dependencies among variables, increasing number of entities in the population and the often limited overlap between data sets\cite{fastLink_paper_2019, review_methods_2014, theoryRL_2018}. The Bayesian graphical entity resolution literature notes that the commonly applied uniform prior on the linkage structure tends to overestimate the number of unique individuals across samples\cite{steorts_bayER_2016, marchant_bayER_2023}, leading to unreliable FDP estimates. Empirical studies further confirm this overestimation of linkage scores\cite{fastLink_paper_2019, flexrlpackage_2024, guha_bay_causal_RL_2022} resulting in an underestimation of the FDP when naively derived from a threshold-based classification.

Despite the increasing use of linked data in research\cite{guha_bay_causal_RL_2022, perined_linkage_2021, opensafely2020, cooneyhealthcare_2023, CoxLinkedData_2024}, there is no universally applicable measure of the error made when linking records. Since linkage errors substantially impact inference\cite{simulrl_2018, regrlinkeddata_2022}, this underscores the need for robust tools to estimate the FDP in RL. To address this gap, we propose a new procedure for estimating the FDP by incorporating synthetic records into the data. The FDP is estimated by assessing whether these records are erroneously linked. This method is inspired by the target-decoy search strategy in proteomics, which estimates false discoveries by comparing sequences with both true target data and generated decoy data\cite{target_decoy_fdr_2012}. Since this approach can be applied to any RL method, it provides a practical and generalisable tool for evaluating linkage error, paving the way for further research to be conducted using reliably linked data.

We illustrate our false discovery estimation procedure using the Perinatal Registry of the Netherlands (PRN), which records births in North Holland between 1999 and 2009. Many have used these data to study pre-term birth, post-term birth, and stillbirth risks by linking siblings records to examine associations between consecutive deliveries\cite{prn0_2012, prn00_2012, prn1_2015, prn2_2022}. In addition, RL has been used to link these data with external sources to investigate more general pregnancy outcomes\cite{Hofprn_2017, prn3_2019}. However, without a unique identifier to validate the linkage, the reliability of inferences drawn from these studies remains uncertain. In this paper, we apply RL to reconstruct siblings pairs and we use our FDP estimation method to select a subset of reliably linked records. As an illustrative example, we estimate the risk of pre-term birth using the linked data, incorporating maternal characteristics and prior deliveries information. Since the true linkage structure is unknown in this context, we further evaluate the proposed procedure using well-established labelled data. We demonstrate how the FDP estimation refines the set of linked records and improves inference results on both simulated and real data from the Survey of Household Income and Wealth (SHIW)\cite{SHIW_BancadItalia_1989}. Additionally, we apply the method to study the evolution of the frailty index in elderly populations using linked data from the National Long Term Care Survey (NLTCS)\cite{NLTCS_NACDA_2006}. 

We introduce the problem and develop our method to estimate the FDP in RL in \Cref{sec 2 problem}. We expand on the issue of which source file to use for synthesis in \Cref{sec 3 good setting} and \Cref{sec 4 good size} and evaluate its scalability and robustness to the key assumptions in \Cref{sec 5 scalability} and \Cref{sec 6 robustness}. We appraise the new estimation procedure across different RL methods and real data sets in \Cref{sec 7 applications}. In the same Section, we highlight the importance of the FDP estimation and its robustness, and demonstrate its role in improving inference reliability. We conclude with guidelines towards improving false discovery in RL.

\section{Problem} \label{sec 2 problem}

Consider the task of linking records across two overlapping data sources originating from the same population, $$\mathcal{A}=\big\{a_i \bigm| i=1,\dots,\nA\big\} \text{ and } \mathcal{B}=\big\{b_j \bigm| j=1,\dots,\nB\},$$ to construct a study population defined by their intersection. Let \smash{$\delta_{i,j}$} indicates whether the pair of records \smash{$(a_i, b_j) \in \mathcal{A} \times \mathcal{B}$} belongs to the same entity (a link: \smash{$\delta_{i,j}=1$}) or to different entities (a non-link: \smash{$\delta_{i,j}=0$}). The RL task therefore targets the set of links: $$\Delta = \big\{ (a_i, b_j) \in \mathcal{A} \times \mathcal{B},\, \delta_{i,j}=1 \big\}.$$ When no unique identifier is available to perfectly determine which pairs of records belong to the same entities, RL procedures must rely on partially identifying variables present in both data sets (e.g.\ birth year, postal code) to link the records.

RL allows to control for confounding variables in inference by expanding the analysis to include more variables, linking longitudinal data for studying long-term outcomes, or exploring secondary outcomes not originally measured. In our task, \smash{$\mathcal B$} is a large file of electronic health records, while \smash{$\mathcal A$} comes from clinical research data. In other examples \smash{$\mathcal B$} represents baseline measurements in a clinical study, while \smash{$\mathcal A$} contains follow-up data with fewer patients or, to recall our main motivation, \smash{$\mathcal B$} gathers first-born babies across a country, while \smash{$\mathcal A$} focuses on subsequent deliveries. Since our interest typically lies in the subsequent analysis, RL is often employed as a tool for data integration rather than as the ultimate aim. Hence the importance of acknowledging and addressing linkage errors, as they may lead to bias in inference.

The RL problem is fundamentally hard. Performance in real world problems depends on the discriminative power of entity features and the dimensions of data sources and their intersection\cite{reflexRL_2020, review_methods_2014, theoryRL_2018}. This calls for an generic procedure for estimating the FDP, which we develop for settings where distinct entities may share similar or identical linkage variables. In addition, we investigate the scalability of the method with respect to the number of entities in the population and the effect of varying degrees of overlap between the data sources.

We assume all records within a data set belong to different underlying entities; we do not tackle the deduplication problem but we show our method's robustness under non-deduplicated data in \Cref{sec 6 robustness}. Moreover we assume that links happen at random i.e.\ the fact that two records from different files relate to the same entity is not related to any observed mechanism; we elaborate further on this assumption in \Cref{distr_link_nonlink} and \Cref{sec 6 robustness}. This setting conveys that the records in both files come from the same distribution, even though registration error mechanisms may differ between the files. It also implies that, within a dataset, the linkage variables of records that form a link have the same distribution as that of records that do not.

\begin{rmk} \label{distr_link_nonlink} 
A central assumption in our methodology is that the distribution of linkage variables is the same for links and non-links---links happen at random---for that it allows us to estimate and sample from the distribution of non-links. This assumption is consistent with the established RL methodologies, the Fellegi-Sunter model and the graphical entity resolution framework. It is particularly realistic when linking variables have low discriminative power, leading to similar distributions for links and non-links\cite{theoryRL_2018}. However, it may not hold when links and non-links differ due to underlying factors (e.g.\ linking a cohort study to a disease registry). This assumption becomes less restrictive in settings where file \smash{$\mathcal{B}$} is large with respect to file $\mathcal{A}$ since most records in \smash{$\mathcal{B}$} do not form a link. Hence, the synthetic records will reflect the distribution of non-links. We show in \Cref{sec 6 robustness} that our method remains robust as we deviate from this assumption.
\end{rmk}

Hereafter, we use \smash{$N$} indexed with a set to denote its size and we index with `synth' any quantity involving a synthetic record in their derivation. By default, when there is no index we refer to (potentially unknown) quantities derived from real records. We use the notation \smash{$\ind$} for the indicator function. We use the term overlap to refer to the intersection between the two data sources and we refer to a record as a set of linkage variables. Finally, without loss of generality, we assume that file \smash{$\mathcal{B}$} contains more observations than file \smash{$\mathcal{A}$} (\smash{$\nA < \nB$}).

\subsection{RL methodology} \label{sec 2 sub 1 rl method}

Several RL methods are available in \texttt{R} and \texttt{Python}. After reviewing recent developments in the field, we focus on three methods that have been specifically developed for the task of linking two unlabelled data sets and have code available in \texttt{R}: \textit{BRL}\cite{sadinle_bay_bipartite_RL_2017}, \textit{FastLink}\cite{fastLink_paper_2019}, and \textit{FlexRL}\cite{flexrlpackage_2024}. \begin{itemize}[topsep=0pt]
    \item \textit{BRL} addresses the limitations of the foundational mixture model proposed by Fellegi and Sunter\cite{FS_RL_1969} by incorporating dependencies among the linkage decisions. This method uses a Gibbs sampler to link records based on binary comparisons of their information. While it provides high-quality results, it requires substantial memory to handle large data sets and, although accessible by inspecting the source code, the package does not explicitly output posterior linkage probabilities.
    \item \textit{FastLink} was developed to mitigate the computational burden of \textit{BRL}. It uses an Expectation-Maximization (EM) algorithm to link the data based on binary comparisons of the information contained in the records.
    \item \textit{FlexRL} avoids the information reduction caused by binary comparisons of records information by modelling the true latent values of the data using an EM algorithm. That way, it can capture more complex relationships between records.
\end{itemize} In addition, we use \textit{SPLink}\cite{splink_2022}, a method for Scalable Probabilistic Linkage developed in \texttt{Python} for the entity resolution task (it thus tackles RL and deduplication), to investigate the scalability of our method with the number of entities in the population and the intertwined effect of varying degrees of overlap in \Cref{sec 5 scalability}, as well as the sensitivity of our method to assumptions deviation in \Cref{sec 6 robustness}. Its implementation is based on that of \textit{FastLink}\cite{fastLink_paper_2019}. Other methods developed in \texttt{Python} tend to rely on older methodologies or require a training set\cite{FEBRL_method_2008, FastHash_method_2017, DeepMatcher_method_2018}. 

The different ways in which probabilistic RL estimates $\Delta$ are comparable; hereupon we refer to it in terms of classification of the linkage scores (or posterior linkage probabilities). Denote by \smash{$d_{i,j}$} the linkage score of a pair of records \smash{$(a_i, b_j)$} obtained with one of the RL methods. Assuming the RL procedure has run long enough for the iterative algorithm to converge, we consider the RL result as fixed (not random). We define a threshold \smash{$\xi \in [0.5,1)$} to construct a set of linked records and estimate \smash{$\Delta$} with: $$D(\xi) = \big\{ (a_i, b_j) \in \mathcal{A} \times \mathcal{B},\, d_{i,j} \in (\xi,1] \big\}.$$ This set must satisfy the one-to-one assignment constraint required for establishing coherent links, which is ensured by setting the lower bound for \smash{$\xi$} at 0.5\cite{tancredi_bay_RL_2011,sadinle_bay_bipartite_RL_2017}. The linkage scores generally follow a bimodal distribution, and the threshold \smash{$\xi$} should separate non-linked records with low probability from linked records with high probability. The quality of the linkage variables influences the modes; more unique values and fewer registration errors make the RL task easier and the separation of linkage scores clearer. The threshold $\xi$, above which records are declared linked, may depend on other parameters of the RL model (like in \textit{BRL}\cite{sadinle_bay_bipartite_RL_2017}) or may be explicitly set (like in \textit{FastLink}\cite{fastLink_paper_2019}, \textit{FlexRL}\cite{flexrlpackage_2024} and \textit{SPLink}\cite{splink_2022}).

We define the real and unknown numbers of true positives (\smash{$TP$}), false positives (\smash{$FP$}), and false negatives (\smash{$FN$}), of the set of linked records as functions of the threshold \smash{$\xi$}: $$TP(\xi) = \sum_{i=1}^{\nA} \sum_{j=1}^{\nB} \ind\{d_{i,j}>\xi\} \ind\{\delta_{i,j}=1\},\:\: FP(\xi) =  \sum_{i=1}^{\nA} \sum_{j=1}^{\nB} \ind\{d_{i,j}>\xi\} \ind\{\delta_{i,j}=0\},\:\:FN(\xi) =  \sum_{i=1}^{\nA} \sum_{j=1}^{\nB} \ind\{d_{i,j}\leq\xi\} \ind\{\delta_{i,j}=1\}.$$ 

We define the False Discovery Proportion (FDP) in the RL task between \smash{$\mathcal{A}$} and \smash{$\mathcal{B}$} to be the (unobserved) proportion of false discoveries among linked pairs as a function of the threshold \smash{$\xi$}: $$\text{FDP}(\xi) = \frac{FP(\xi)}{TP(\xi) + FP(\xi)}.$$ The naive probabilistic FDP estimate evoked in the introduction and derived from the RL modelling is defined\cite{fastLink_paper_2019, fdr_genomewide_2003} as follows: \begin{align} \label{probFDP}
    \text{prob}\widehat{\text{FDP}}(\xi) = \frac{ \sum_{i=1}^{\nA} \sum_{j=1}^{\nB} \ind \big\{ d_{i,j} > \xi \big\} (1-d_{i,j}) }{ \sum_{i=1}^{\nA} \sum_{j=1}^{\nB} \ind \big\{ d_{i,j} > \xi \big\} }.
\end{align}

\subsection{Estimation procedure} \label{sec 2 sub 2 procedure}

To estimate the FDP for the RL task linking \smash{$\mathcal{A}$} and \smash{$\mathcal{B}$}, we propose augmenting file \smash{$\mathcal{B}$} by sampling \smash{$\nS$} synthetic records from the estimated empirical distribution of the data contained in \smash{$\mathcal{B}$}. For that, we use a synthesiser as we explain further on in \Cref{sec 2 sub 3 synthesisers}. We denote \smash{$\tilde{\mathcal{B}}$} the concatenation of file \smash{$\mathcal{B}$} and the records synthesised from its estimated empirical distribution. By linking file \smash{$\mathcal{A}$} and the augmented file \smash{$\tilde{\mathcal{B}}$} with an RL algorithm, we can estimate the amount of falsely linked records thanks to the pairs formed with a synthetic record. This allows us to construct an estimator for the FDP.

We define the numbers of synthetic false positives and real linked records as functions of the threshold \smash{$\xi$}: $${FP}_{\text{synth}}(\xi) = \sum_{i=1 \vphantom{\nB} }^{\nA \vphantom{\nS}} \sum_{j=\nB+1}^{\nB+\nS} \ind\{d_{i,j} > \xi\},\:\: N_{\text{real linked}}(\xi) = \sum_{i=1}^{\nA} \sum_{j=1}^{\nB} \ind\{ d_{i,j} > \xi\} = TP(\xi) + FP(\xi).$$ Since we assume that the iterative RL process has converged, it does not contribute to the randomness of the results. So all randomness originates from the synthesiser. Henceforth, \smash{$\mathbb E_{\text{synth}} [ {FP}_{\text{synth}}(\xi) ]$} represents the average number of synthetic false positives resulting from synthesising data.

When running RL between file \smash{$\mathcal{A}$} and the augmented file \smash{$\tilde{\mathcal{B}}$} we differentiate real linked pairs from synthetic false positives in the linkage results, hence we then estimate the FDP as: \begin{align} \label{hatFDP}
\hatFDP(\xi) = \frac{{FP}_{\text{synth}}(\xi) \cdot \frac{\nB}{\nS}}{N_{\text{real linked}}(\xi)}.
\end{align} 

As the synthetic records do not correspond to real individuals, we use the RL task linking real and augmented data sets \smash{$\mathcal{A}$} and \smash{$\tilde{\mathcal{B}}$} to estimate the false discoveries among real data when linking \smash{$\mathcal{A}$} and \smash{$\mathcal{B}$}. Though, since synthetic records may resemble records that form a link, we need to ensure that the augmented \smash{$\tilde{\mathcal{B}}$} corresponds to a (theoretical) version of \smash{$\mathcal{B}$} only augmented with non-links, to legitimately label synthetic linked pairs as false. We denote \smash{$\mathcal{B}^\prime$} this theoretical file composed of records from \smash{$\mathcal{B}$} and synthetic non-links. The procedure we propose thus generates an unbiased estimator under the following conditions:\begin{itemize}[topsep=0pt]
    \item[\hypertarget{req1}{$(i)$}] Synthetic records represent records from \smash{$\mathcal{B}$} that do not form a link in $\mathcal{A}$,
    \item[\hypertarget{req2}{$(ii)$}] Synthetic records minimally affect the RL process.
\end{itemize}

\paragraph*{Condition $(i)$ induces \smash{$\mathbb{E}[\hatFDP_{\texttt{RL}(\mathcal{A},\tilde{\mathcal{B}})}(\xi)] = \text{FDP}_{\texttt{RL}(\mathcal{A},\mathcal{B}^\prime)}(\xi)$}}

The first condition defines pairs involving a synthetic record as non-links (\smash{$TN$} and \smash{$FP$}). Note that these records may copy records from \smash{$\mathcal{B}$} carrying information that is likely to be sampled, reflecting the low discriminative power of linkage variables. Thus, for a record in \smash{$\mathcal{A}$} indexed by \smash{$i \in \{1, \dots, \nA\}$} and a synthetic record in \smash{$\mathcal{B}$} indexed by \smash{$j \in \{\nB+1, \dots, \nB+\nS\}$} we set \smash{$\delta_{i,j}=0$}. It follows that there should not be more synthetic linked pairs than real linked pairs (in proportion) since the synthetic records should not be more similar to records in \smash{$\mathcal{A}$} than are the real records in \smash{$\mathcal{B}$}: \begin{align} \label{domain ass eq}
    \frac{\mathbb E_{\text{synth}} [{FP}_{\text{synth}}(\xi)]}{\nS} \leq \frac{N_{\text{real linked}}(\xi)}{\nB}.
\end{align} Under \hyperlink{req1}{$(i)$} the proportions of synthetic \smash{$FP$} relative to all synthetic pairs and of real \smash{$FP$} to all real pairs should be equal: \begin{align} \label{synth ass eq}
    \frac{\mathbb E_{\text{synth}} [{FP}_{\text{synth}}(\xi)]}{\nA \cdot \nS} = \frac{FP(\xi)}{\nA \cdot \nB}.
\end{align}

With \hyperlink{req1}{$(i)$}, we implicitly assume that we are able to sample from the distribution of non-links. It is the case when links happen at random, as all records are thus assumed to be independent samples from the population linkage variables, whose distribution can be estimated. One can also achieve this by estimating the empirical data distribution on provided negative controls (records we are sure should not link). Then, we can apply a resampling strategy to generate records that will legitimately be non-links. In practice, we estimate the empirical distribution of the population linkage variables with a synthesiser, and sample synthetic records. We mention solutions to evaluate the synthetic data quality in \Cref{sec 2 sub 3 synthesisers}.

Whether we may assume that links happen at random or instead can use reliable negative controls to estimate the targeted distribution should be assessed per application. Although \cref{synth ass eq}---which follows from condition \hyperlink{req1}{$(i)$} and ensures accuracy of our estimator---cannot be verified with unlabelled data, bias can in some cases be detected via \cref{domain ass eq} which is equivalent to \smash{$\mathbb E_{\text{synth}} [\hatFDP(\xi)] \leq 1$}. Indeed, since \cref{synth ass eq} implies \cref{domain ass eq}, by contraposition an estimate exceeding one would indicate the presence of bias.

\paragraph*{Condition 
$(ii)$ induces \smash{$\text{FDP}_{\texttt{RL}(\mathcal{A}, \mathcal{B}^\prime)}(\xi) = \text{FDP}_{\texttt{RL}(\mathcal{A}, \mathcal{B})}(\xi)$}}

The second condition ensures that the FDP on real data when linking \smash{$\mathcal{A}$} and a properly augmented \smash{$\mathcal{B}^\prime$} is actually equal to the FDP on real data when linking \smash{$\mathcal{A}$} and \smash{$\mathcal{B}$}. This is important because the FDP usually increases as the overlap between data sets decreases, which would be enforced by augmenting file \smash{$\mathcal{B}$}. Requirement \hyperlink{req2}{$(ii)$} ensues if we synthesise records in the right setting. The number of synthetic observations \smash{$\nS$} must be carefully set to ensure accurate estimation of the FDP while minimising the impact of synthetic records on the original RL task. We investigate the population from which we should sample and the size of the synthetic set in \Cref{sec 3 good setting} and \Cref{sec 4 good size}.

Given that these conditions hold, it is straightforward to see that \smash{$\hatFDP(\xi)$} is unbiased for the FDP on real data: \begin{equation*}
\mathbb E_{\text{synth}} [ \hatFDP(\xi) ] =  \frac{\mathbb E_{\text{synth}} [ {FP}_{\text{synth}}(\xi) ] \cdot \frac{\nB}{\nS}}{N_{\text{real linked}}(\xi)} \overset{\cref{synth ass eq}}{=} \frac{FP(\xi)}{N_{\text{real linked}}(\xi)} = \frac{FP(\xi)}{TP(\xi)+FP(\xi)}.
\end{equation*}

Although the choice of a formula in \cref{hatFDP} seems natural, an alternative estimate was sometimes used in the literature\cite{target_decoy_fdr_2012, target_decoy_alternative_2007}: \begin{align} \label{hatFDPS}
\hatFDPS(\xi) &= \frac{{FP}_{\text{synth}}(\xi) \cdot (1+\frac{\nB}{\nS})}{N_{\text{all linked}}(\xi)}.
\end{align} This estimate accounts for all falsely linked pairs (\smash{$FP_{\text{synth}}$} and the estimation of real \smash{$FP$}) in the numerator and all linked pairs in the denominator. It calculates the FDP on augmented data where, $$N_{\text{all linked}}(\xi) = \sum_{i=1}^{\nA \vphantom{\nS} } \sum_{j=1}^{\nB+\nS} \ind\{ d_{i,j} > \xi\} = TP(\xi) + FP(\xi) + FP_{\text{synth}}(\xi),$$ which results in a biased estimate of FDP on real data as \cref{hatFDPS} focuses on the actual task on augmented data. Despite a former interest in this formula, which accounts for the impact of the estimation method on the RL method, we argue that synthetic pairs should not be included in the estimate, as concluded in the target-decoy approaches\cite{target_decoy_fdr_2012}. 

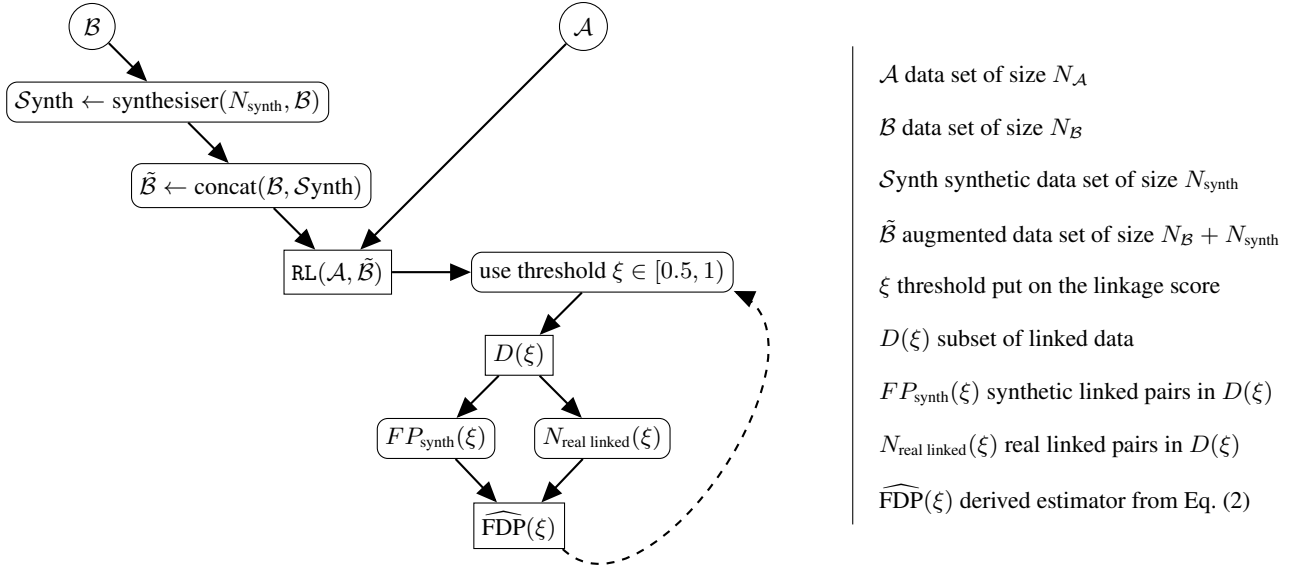
\begin{figure}
    \raggedright
    \begin{tikzpicture}[
      every node/.style={align=center},
      assumption/.style={draw, circle},
      check/.style={draw, rounded corners},
      result/.style={draw},
      arrow/.style={->, thick}
    ]
    \node[assumption] (B) at (-3.25,3.25) {$\mathcal{B}$};
    \node[check] (synth) at (-2.25,2.25) {$\mathcal{S}\text{ynth} \gets \text{synthesiser}(\nS,\mathcal{B})$};
    \node[assumption] at (3.25,3.25) (A) {$\mathcal{A}$};
    \node[check] (tildeB) at (-1.145,1.145){$\tilde{\mathcal{B}} \gets \text{concat}(\mathcal{B}, \mathcal{S}\text{ynth})$};
    \node[result] (rl) at (0,0){$\texttt{RL}(\mathcal{A},\tilde{\mathcal{B}})$};
    \node[check] (loop) at (3.5,0) {use threshold $\xi \in [0.5,1)$};
    \node[result] (setxi) at (2.4,-1.1) {$D(\xi)$};
    \node[check] at (1.3,-2.2) (fpsynth) {$FP_{\text{synth}}(\xi)$};
    \node[check] at (3.5,-2.2) (Nlinked) {$N_{\text{real linked\vphantom{y}}}(\xi)$};
    \node[result] at (2.4,-3.355) (fdp) {$\hatFDP(\xi)$};
    \draw[arrow] (B) -- (synth);
    \draw[arrow] (synth) -- (tildeB);
    \draw[arrow] (tildeB) -- (rl);
    \draw[arrow] (A) -- (rl);
    \draw[arrow] (rl) -- (loop);
    \draw[arrow] (loop) -- (setxi);
    \draw[arrow] (setxi) -- (fpsynth);
    \draw[arrow] (setxi) -- (Nlinked);
    \draw[arrow] (fpsynth) -- (fdp);
    \draw[arrow] (Nlinked) -- (fdp);
    \draw[arrow, dashed] (fdp) to[bend right=100] (loop);
    \node[right] (top) at (6.7,3.1) {};
    \node[right] (1) at (6.7,2.6) {\hspace{10pt}$\mathcal{A}$ data set of size $\nA$};
    \node[right] (2) at (6.7,1.9) {\hspace{10pt}$\mathcal{B}$ data set of size $\nB$};
    \node[right] (3) at (6.7,1.2) {\hspace{10pt}$\mathcal{S}\text{ynth}$ synthetic data set of size $\nS$};
    \node[right] (4) at (6.7,0.5) {\hspace{10pt}$\tilde{\mathcal{B}}$ augmented data set of size $\nB + \nS$};
    \node[right] (4) at (6.7,-0.2) {\hspace{10pt}$\xi$ threshold put on the linkage score};
    \node[right] (4) at (6.7,-0.9) {\hspace{10pt}$D(\xi)$ subset of linked data};
    \node[right] (4) at (6.7,-1.6) {\hspace{10pt}$FP_{\text{synth}}(\xi)$ synthetic linked pairs in $D(\xi)$};
    \node[right] (4) at (6.7,-2.3) {\hspace{10pt}$N_{\text{real linked\vphantom{y}}}(\xi)$ real linked pairs in $D(\xi)$};
    \node[right] (4) at (6.7,-3) {\hspace{10pt}$\hatFDP(\xi)$ derived estimator from \cref{hatFDP}};
    \node[right] (bottom) at (6.7,-3.45) {};
    \draw[] (top) -- (bottom);
    \end{tikzpicture}
    \caption{\raggedright Flow chart summarising the false discovery estimation procedure in RL.}
    \label{fig:flowchartalgo}
\end{figure}

We summarise our estimation procedure in \Cref{algo:fdp} and \Cref{fig:flowchartalgo}.

\begin{algorithm}
\caption{FDP estimation in RL}\label{algo:fdp}
\begin{algorithmic}
\Require RL algorithm, synthesiser, file $\mathcal{A}$, file $\mathcal{B}$, $\nA < \nB$
\State $\mathcal{S}\text{ynth} \gets \text{synthesiser}(\nS,\mathcal{B})$ \Comment{Synthesise $\nS=0.10 \times \nB$ records based on file $\mathcal{B}$}
\State $\tilde{\mathcal{B}} \gets \text{concat}(\mathcal{B}, \mathcal{S}\text{ynth})$
\State $\big\{ (a_i,b_j,p),\, a_i \in \mathcal{A},\, b_j \in \tilde{\mathcal{B}},\, p \in [0,1] \big\} \gets \text{RL}( \mathcal{A}, \tilde{\mathcal{B}})$ \Comment{Run RL between file $\mathcal{A}$ and augmented file $\tilde{\mathcal{B}}$}
\For{$\xi \in [0.5,1)$}
\State $D(\xi) \gets \big\{ (a_i,b_j,p),\, a_i \in \mathcal{A},\, b_j \in \tilde{\mathcal{B}},\, p \in (\xi,1] \big\}$
\State $FP_{\text{synth}}(\xi) \gets \sum_{\ell \in D(\xi)} \ind{\{ \ell \coloneqq (a_i,b_j,p) \in D(\xi),\, b_j \in \mathcal{S}\text{ynth} \}}$
\State $N_{\text{real linked}}(\xi) \gets \sum_{\ell \in D(\xi)} \ind{\{ \ell \coloneqq (a_i,b_j,p) \in D(\xi),\, b_j \in \mathcal{B} \}}$
\State $\hatFDP(\xi) \gets \frac{FP_{\text{synth}}(\xi) \cdot \nB/\nS}{N_{\text{real linked}}(\xi)}$
\EndFor
\Ensure Sets $\big\{ (a_i,b_j,p),\, a_i \!\in\! \mathcal{A},\, b_j \!\in\! \mathcal{B},\, p \!\in\! (\xi,1] \big\}_{\xi \in [0.5,1)}$ of real linked records and corresponding $\hatFDP(\xi)$
\end{algorithmic}
\end{algorithm}

With this approach, our aim is to give researchers a flavour of the FDP associated with linked data. By applying the proposed methodology one can obtain a sequence of subset of linked data jointly with the corresponding estimate of the FDP. The higher the threshold \smash{$\xi \in [0.5,1)$} used to define a subset of linked data, the more stringent the linked set and therefore the lower the FDP and the estimated FDP. One can plot for instance the evolution of the estimated FDP as well as the evolution of the number of linked pairs as functions of the threshold. The minimal threshold one should take is 0.5 in order to ensure coherent links (fulfilling the one-to-one assignment constraint). One will have to find the balance between a desirable estimated FDP and an adequate number of observations for the inference; this decision depends on each application. The maximal threshold may not lead to a satisfactory FDP in which case it becomes evident that the linkage is not reliable.

For a given threshold, one should run multiple procedures to properly estimate \smash{$\mathbb E_{\text{synth}} [ {FP}_{\text{synth}}(\xi) ]$} and obtain, for a given threshold, a point estimate for the FDP and its standard error. Such an aggregated estimate of the FDP can be derived either from a truncated average on the estimates upper bounded by \smash{$1$}, or by taking the minimum between the estimates and $1$ and averaging it, or else by taking the median over the set of estimates. In the real data applications of \Cref{sec 7 applications}, we only observed some estimates exceeding \smash{$1$} in small sample size contexts.

To build the above procedure, we explored the possible choices of a formula for the FDP estimate (aforementioned) and the entangled properties required from the synthesiser. Multiple methods have been developed to generate synthetic data; we present two methods in the following \Cref{sec 2 sub 3 synthesisers}. We chose to augment file \smash{$\mathcal{B}$} with \smash{$\nS=0.10\times\nB$} records because this configuration minimises the impact of synthetic records on the RL process. We elaborate on this aspect in \Cref{sec 3 good setting}, where we investigate potential synthetic database constructions, and in \Cref{sec 4 good size}, where we evaluate the impact of data sets size on the estimation. The reliability of the FDP estimate depends on all these choices. We evaluate the scalability and robustness of our method in \Cref{sec 5 scalability} and \Cref{sec 6 robustness}. We illustrate the efficacy and the value of estimating the FDP in applications in \Cref{sec 7 applications}.

\subsection{Data synthesisers} \label{sec 2 sub 3 synthesisers}

After reviewing the different possibilities to generate synthetic records (with categorical variables), based on software availability on \texttt{R} and \texttt{Python} and computational feasibility, we selected \textit{synthpop}\cite{synthpop_2016} and \textit{arf}\cite{arf_2023}.\begin{itemize}[topsep=0pt]
    \item \textit{Synthpop} uses sequential modelling to generate each column of the synthetic data from conditional distributions fitted to the original data using classification and regression trees. The first column is generated by random sampling; therefore, we often set as the first column the variable with the largest number of unique values. This method may be slow in practice.
    \item \textit{Arf} uses generative modelling, namely adversarial random forest, to synthesise data from the estimated density of the original data. Trees gradually learn the data properties by training the classification of data into real or synthetic. This method is fast in the low dimensional context of few linkage variables and takes advantage of large data sets.
\end{itemize} Both methods have been shown to perform similarly in terms of Negative Log-Likelihood. Other possibilities are provided on \texttt{Python}: \textit{gretel-synthetics}, \textit{nbsynthetic}, \textit{DataSynthesizer}, but we do not explore them.

RL often relies on categorical variables which may make the synthesis very slow, especially when these are of high cardinality. Examples of linkage variables are birth year, postal code, socioeconomic status, sex. To ease synthesis, we suggest grouping values of high cardinality variables into higher level categories. While lowering their identifying strength, it encompasses the limitations encountered by the synthesiser. In contexts where names are available (rare in medical data where our primary interest lies), phonetic groups may be formed by converting string data into numerical values using soundex code\cite{RusselSoundexCode1, RusselSoundexCode2}. For data covering a country (lots of postal code or addresses) or a long period of time (lots of birth years), one may need to generate those variables as continuous. Addresses, similarly to postal codes, may be grouped hierarchically by municipality and neighbourhood. They may then be interpreted as continuous, assuming an underlying geographical smoothness of the conditional distribution of other variables with respect to it (nearby locations tend to be more similar in demographic variables), and similarly for birth year.

A good data synthesis is crucial for our methodology. Rare categories, data points located on the edge of joint distributions, or complex relationships between variables may be difficult to synthesise. Nevertheless, since most linkage variables are categorical, the task reduces to sampling from multinomial distributions. Flexible data synthesisers often do not rely on strong parametric assumptions about the data and are well suited for this task; they can capture edge and rare cases and they can model dependencies among linkage variables.

For our method to be valid, we need to ensure that the synthetic data resemble the non-links. With unlabelled data, this is not verifiable without further assumption. Thus, we assumed so far that links happen at random, and we show in \Cref{sec 6 robustness} that our method is fairly robust against deviations from this assumption in various realistic settings. Ultimately, we need to certify that the synthetic data resemble the initial records (values frequencies and joint variables structures should be preserved). Multiple explainable AI techniques may be used to evaluate synthetic tabular data\cite{kapar_xai_2025}. In order to confirm the quality of the data synthesised, one may train a classifier to distinguish real data from synthetic data and use the AUC as a primary tool for marginal explanations; we do so in \Cref{sec 5 scalability} and \Cref{sec 6 robustness}. In addition, global and local tools---such as feature importance measures, feature effect plots, Shapley values and counterfactuals---may be used for conditional explanations to understand how well variables were generated over their support and how the synthetic data may differ from the real data\cite{kapar_xai_2025}.

\section{Source for synthetic data}\label{sec 3 good setting}

To estimate false positives with synthetic linked pairs, we could synthesise records in \smash{$\mathcal A$} or in \smash{$\mathcal B$}. Falsely linked pairs usually are formed of records with nearly identical information, and form legitimate ambiguities in the linkage. This is mostly due to the few number of unique values in the linkage variables used in RL, which makes the task non-trivial. We investigated three procedures to generate synthetic data that represent non-links. \begin{itemize}[topsep=0pt]
    \item[\hypertarget{proa}{$(a)$}] Synthesise records based on one file (either \smash{$\mathcal A$} or \smash{$\mathcal B$}), concatenate them with the source file, and run RL between the augmented file and the other source.
    \item[\hypertarget{prob}{$(b)$}] Synthesise records from both data sources and run RL between the two augmented files.
    \item[\hypertarget{proc}{$(c)$}] Generate synthetic sets from both \smash{$\mathcal A$} and \smash{$\mathcal B$} and run RL between the synthetic files.
\end{itemize}

Procedures \smash{$(b)$} and \smash{$(c)$} potentially have some undesirable properties, which may render a violation of the aforementioned requirements to obtain an unbiased estimator. When synthesising records from both files (and running RL on the synthetic sets either concatenated with real data or separately), sampling from the distribution of records that do not form a link becomes more restrictive. Particularly, we need to assume some structure on file \smash{$\mathcal{A}$} unlikely to hold in the setting we tackle where \smash{$\mathcal{A}$} is a smaller file containing more specific data. In such context we have fewer records not forming a link to learn the targeted distribution. Moreover, we are more likely to affect the RL process due to the uncertainty introduced by more synthetic records. Indeed, synthetic pairs would be of several forms: a synthetic from \smash{$\mathcal{A}$} (respectively \smash{$\mathcal{B}$}) with a record in \smash{$\mathcal{B}$} (respectively \smash{$\mathcal{A}$}), or two synthetic records. When two synthetic records carry identical information, labelling the pair as either true or false becomes ambiguous. In addition, we have no interest in comparing synthetic to synthetic since we do not target matching patterns but rather the confusion any RL method faces when selecting a pool of records in \smash{$\mathcal{B}$}, all looking alike and matching the information contained in a given record from \smash{$\mathcal{A}$}.

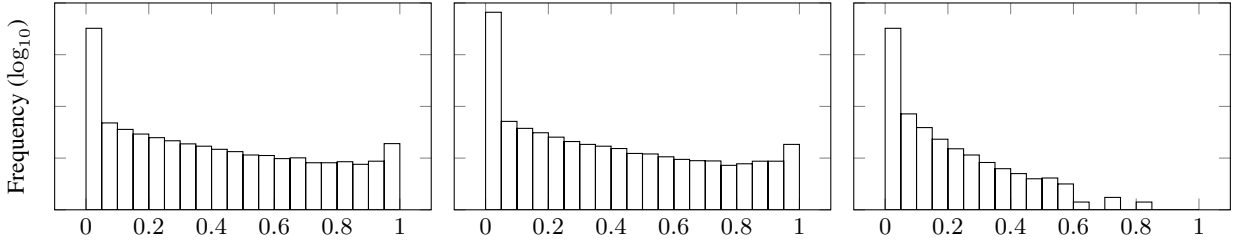
\begin{figure}
    \centering
    \begin{tikzpicture}
    \pgfplotsset{
    width=0.38\textwidth,
    height=0.25\textwidth}
    \begin{axis}[
        ymin=0, ymax=8,
        ytick={0,2,4,6,8},
        yticklabels={,,,,},
        ylabel={Frequency ($\log_{10}$)}
        ]
    \addplot+[ybar interval, mark=no, draw=black
    ] plot coordinates { (0, 7.02) (0.05, 3.36) (0.1, 3.11) (0.15, 2.93) (0.2, 2.79) (0.25, 2.67) (0.3, 2.55) (0.35, 2.46) (0.4, 2.34) (0.45, 2.25) (0.5, 2.12) (0.55, 2.10) (0.6, 1.98) (0.65, 2.01) (0.7, 1.82) (0.75, 1.82) (0.8, 1.86) (0.85, 1.76) (0.9, 1.88) (0.95, 2.56) (1, 2.56) };
    \end{axis}
    \end{tikzpicture}
    \begin{tikzpicture}
    \pgfplotsset{
    width=0.38\textwidth,
    height=0.25\textwidth}
    \begin{axis}[
        ymin=0, ymax=8,
        ytick={0,2,4,6,8},
        yticklabels={, , , , } 
        ]
    \addplot+[ybar interval, mark=no, draw=black
    ] plot coordinates { (0, 7.64) (0.05, 3.42) (0.1, 3.15) (0.15, 2.98) (0.2, 2.81) (0.25, 2.64) (0.3, 2.53) (0.35, 2.46) (0.4, 2.37) (0.45, 2.18) (0.5, 2.16) (0.55, 2.05) (0.6, 1.95) (0.65, 1.90) (0.7, 1.89) (0.75, 1.72) (0.8, 1.78) (0.85, 1.88) (0.9, 1.88) (0.95, 2.53) (1, 2.53) };
    \end{axis}
    \end{tikzpicture}
    \begin{tikzpicture}
    \pgfplotsset{
    width=0.38\textwidth,
    height=0.25\textwidth}
    \begin{axis}[
        ymin=0, ymax=8,
        ytick={0,2,4,6,8},
        yticklabels={, , , , } 
        ]
    \addplot+[ybar interval, mark=no, draw=black
    ] plot coordinates { (0, 7.02) (0.05, 3.71) (0.1, 3.18) (0.15, 2.73) (0.2, 2.36) (0.25, 2.12) (0.3, 1.83) (0.35, 1.59) (0.4, 1.40) (0.45, 1.20) (0.5, 1.23) (0.55, 1) (0.6, 0.30) (0.65, 0.00) (0.7, 0.48) (0.75, 0.00) (0.8, 0.30) (0.85, 0.00) (0.9, 0.00) (0.95, 0.00) (1, 0.00) };
    \end{axis}
    \end{tikzpicture}
    \caption{Linkage scores obtained from \textit{FlexRL} on the SHIW data, Centre of Italy (3,450 and 3,046 records with 45\% of overlap). From left to right, when using only real records: \texttt{RL}($\mathcal{A}, \mathcal{B}$), when using procedure \protect\hyperlink{proa}{$(a)$}: \texttt{RL}($\mathcal{A}, \tilde{\mathcal{B}}$) and when using procedures \protect\hyperlink{prob}{$(b)$} or \protect\hyperlink{proc}{$(c)$}. The bimodal distribution is representing the mixture between linked and non-linked.}
    \label{fig-SHIWprobaDeltaHist}
\end{figure}

In practice, in the SHIW and the NLTCS applications, we observed that procedures \smash{$(b)$} and \smash{$(c)$} create too many records carrying identical linkage variables to some real records, introducing too much uncertainty in the RL processes tested. \textit{BRL} tends to avoid linking records when the data sources become larger or when the noise coming from synthetic records is excessive. \textit{FastLink} links as many pairs between two fully synthetic sets (\smash{$FP_{\text{synth}}$}) as it does between two real data sources (\smash{$FP+TP$}), whereas \smash{$FP_{\text{synth}}$} should estimate \smash{$FP$}. This aligns with findings from the original paper, noticing larger error rates with smaller proportion of links\cite{fastLink_paper_2019}. Finally, \textit{FlexRL} fails to build the expected bimodal distribution of linkage scores in presence of many synthetic records. \Cref{fig-SHIWprobaDeltaHist} shows the distributions of the linkage scores in the proposed procedure using augmented file \smash{$\tilde{\mathcal{B}}$} and in the original RL task between \smash{$\mathcal{A}$} and \smash{$\mathcal{B}$}. We observe that the linkage scores are not affected by procedure \smash{$(a)$}, i.e.\ \texttt{RL}(\smash{$\mathcal{A}$,$\tilde{\mathcal{B}}$}) and \texttt{RL}(\smash{$\mathcal{A}$,$\mathcal{B}$}) yield similar outputs. We make the same observation on the linkage scores of \smash{$TP$} and real \smash{$FP$}. In addition we show the distribution of the linkage scores when augmenting both files like in procedure \smash{$(b)$} or \smash{$(c)$} to be damaged by the excessive uncertainty introduced by the amount of synthetic pairs. We therefore rule out these methods and synthesise records either in \smash{$\mathcal A$} or in \smash{$\mathcal B$}, estimating the false positives with pairs involving one synthetic record.

In the RL task we depicted, file \smash{$\mathcal{A}$} should be seen as the reference and, for each of its records, one should look for the most coherent link to build in file \smash{$\mathcal{B}$} (if any). Generating records from source \smash{$\mathcal{A}$} and using the concatenation of real and synthetic records as the reference file in the RL process thus poses philosophical challenges. For practical matters, it is also convenient that \smash{$\mathcal{A}$} remains the smallest file. Moreover, the synthesiser will better learn the multivariate distribution of the data from the largest source. We therefore argue that it is more sensible to synthesise records in \smash{$\mathcal{B}$} (the largest file) and estimate the \smash{$FP$} with pairs linking a real record from \smash{$\mathcal{A}$} with a synthetic record in \smash{$\mathcal{B}$}. Henceforth, we run RL between file \smash{$\mathcal{A}$} and the augmented file \smash{$\tilde{\mathcal{B}}$} made of its original records as well as synthetic records generated from the same distribution.

\section{Impact of the size of the synthetic set}\label{sec 4 good size}

The performance of RL depends on the size of the sets to be linked. Indeed, for a given record in \smash{$\mathcal{A}$} with the same linkage variables, there are more potential records in \smash{$\mathcal{B}$} to form a link with when the files are larger. Ideally, to minimally alter the results from RL by adding synthetic data, a single synthetic record should be generated in file \smash{$\mathcal{B}$}. By repeating the procedure many times, one could obtain an estimate of the number of false positives with the counts of runs where the synthetic record was linked. However, since most modern RL methods are computationally expensive, this approach is impractical.

Because of time and resources constraints, RL is often applied with blocking strategies\cite{sadinle_bay_bipartite_RL_2017, dblink_2021, blocking_methods_2014}, i.e.\ the data sources are partitioned into blocks based on a variable (e.g.\ same postal code, same decade of birth) which is supposed to be free from registration errors. Notwithstanding the inherent restriction of blocking, RL is often more accurate on such smaller scale. It is therefore necessary to investigate the estimation of the FDP with and without blocking. Moreover, since we expect the FDP estimation procedure to be affected by the scale of the sources to be linked, we explored the appropriate size of the synthetic data set, as a fraction of \smash{$\nB$}: \smash{$\nS = \alpha \times \nB$}, with \smash{$\alpha \in (0,0.20]$}.

\begin{figure}[!ht]
  \centering
  \begin{tikzpicture}[x=1pt,y=0.5pt]
  \pgfplotsset{
    width=0.5\textwidth,
    height=0.34\textwidth}
    \begin{axis}[
      legend entries={\textit{BRL}, \textit{FastLink}, \textit{FlexRL}},
      xmin=0, xmax=7,
      ymin=0, ymax=0.3,
      axis lines=middle,
      ytick={-0.01,0,0.01,0.02,0.03,0.04,0.05,0.06,0.07,0.08,0.09,0.10,0.11,0.12,0.13,0.14,0.15,0.16,0.17,0.18,0.19,0.20,0.21,0.22,0.23,0.24,0.25,0.26,0.27,0.28,0.29,0.3},
      xtick={0,1,2,3,4,5,6},
      xticklabels={0,1\%,2\%,5\%,10\%,15\%,20\%},
      yticklabels={,0,,,,,0.05,,,,,0.10,,,,,0.15,,,,,0.20,,,,,0.25,,,,,0.3},
      xlabel style={at={(1.05,0)},below right},
      xlabel={$\nS$ as $\%$ of $\nB$},
      ylabel style={at={(-0.15,0.95)},above left},
      ylabel={RMSE},
      legend style={draw=none, at={(1.25,1)}},
      legend cell align={left}
      ]
      \addplot [
            mark=*,
            error bars/.cd,
                y dir=both,y explicit,
        ] table [x=x,y=y,y error=err] {
            x y err
            1 0.198 0 \\
            2 0.213 0 \\
            3 0.142 0 \\
            4 0.114 0 \\
            5 0.076 0 \\
            6 0.065 0 \\
        };
      \addplot [
            mark=square*,
            mark options={draw=black,fill=white},
            error bars/.cd,
                y dir=both,y explicit,
        ] table [x=x,y=y,y error=err] {
            x y err
            1 0.192 0 \\
            2 0.091 0 \\
            3 0.089 0 \\
            4 0.061 0 \\
            5 0.057 0 \\
            6 0.048 0 \\
        };
      \addplot [
            mark=*,
            mark options={draw=black,fill=white},
            error bars/.cd,
                y dir=both,y explicit,
        ] table [x=x,y=y,y error=err] {
            x y err
            1 0.290 0 \\
            2 0.177 0 \\
            3 0.147 0 \\
            4 0.118 0 \\
            5 0.125 0 \\
            6 0.104 0 \\
        };
    \end{axis}
  \end{tikzpicture}
  \caption{RMSE of the FDP estimates over 10 runs of the procedure on the SHIW data, Centre of Italy (3,450 and 3,046 records with 45\% of overlap). The true FDP decreases from 0.27 to 0.23 for \textit{BRL} and from 0.43 to 0.40 for \textit{FlexRL}, while it remained at 0.73 for \textit{FastLink}. The sensitivity decreases from 0.16 to 0.1 for \textit{BRL}, from 0.19 to 0.15 for \textit{FastLink} and from 0.43 to 0.40 for \textit{FlexRL}. We observed the same behaviour on other areas of the SHIW data and on the NLTCS data.}
  \label{fig-impact size synthetic}
\end{figure}
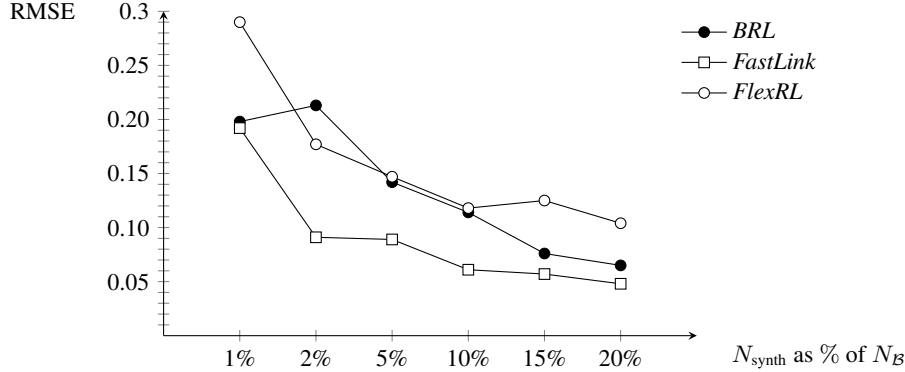

Across datasets of all scales (both with and without blocking), we observed larger variability in the bias of the FDP estimate when generating 1\% to 5\% of \smash{$\nB$} synthetic records (\smash{$\nS \leq 0.05 \times \nB$}). However, this variability decreases as \smash{$\nS$} increases. In RL applied to both small and large data sets, with and without blocking, the FDP estimate stabilises when generating at least 10\% of \smash{$\nB$} synthetic records(\smash{$\nS \geq 0.1 \times \nB$}). We illustrate the evolution of the root mean square error combining bias and variance as a function of \smash{$\nS$} using a subset of the SHIW data in \Cref{fig-impact size synthetic}.

We initially expected that estimating the FDP in RL by adding synthetic records to the data would negatively affect the RL process as it would hinder the task. However, we do not observe such impact. Instead, we note that as \smash{$\nS$} increases---thereby introducing more noise---RL methods become more conservative: the true FDP decreases, but the sensitivity also declines; fewer links are detected and fewer errors are made.

As we observe a levelling off from \smash{$10\%$} onwards in \Cref{fig-impact size synthetic}, in the applications we generate \smash{$\nS= 10\% \times \nB$} synthetic records which we concatenate to file \smash{$\mathcal{B}$}. We apply RL to file $\mathcal{A}$ and the augmented file \smash{$\tilde{\mathcal{B}}$}. However, the number of synthetic records required for a stable FDP estimation may vary depending on factors such as the overlap between data sets and the discriminative power of linkage variables. While adding more synthetic records can reduce the estimator variance, it also increases the computational effort and may alter the behaviour of the RL algorithm, making it more conservative. The size of the synthetic set should balance these aspects.

\section{Scalability}\label{sec 5 scalability}

RL modelling is sensitive to the low discriminative power of linkage variables, to the increasing number of entities in the data sets and to the often limited overlap between data sets, not to mention registration errors and dependencies among linkage variables\cite{review_methods_2014}. In this section, we illustrate the scalability of the method with respect to the underlying number of entities in the data sets. To wit, we use a method for Scalable Probabilistic Linkage developed in \texttt{Python}: \textit{SPLink}\cite{splink_2022} which allows us to show the scalability of our estimation procedure on very large data files.

\begin{table}[ht]
\centering
\begin{tabular}{|c|c|c|c|c|}
\hline
\multicolumn{3}{|c|}{ } & overlap 0.35 & overlap 0.75  \\
\hline 
\multirow{12}{*}{\rotatebox{90}{$\nA << \nB$}} & \multirow{5}{*}{Discr. level 0.85} & $\text{FDP}$ & 0.604 (0.001) & 0.416 (0.001)  \\
 & & $\mathbf{\widehat{\textbf{FDP}}}$ bias & 0.003 (0.012) & 0.073 (0.007)  \\
 & & $\text{prob}\widehat{\text{FDP}}$ bias & -0.409 (0.013) & -0.313 (0.011) \\
 & & condition \ref{synth ass eq} & $e^{-07}$ ($e^{-07}$) & $e^{-07}$ ($e^{-08}$) \\
 & & AUC synth & 0.515 (0.003) & 0.518 (0.003) \\
\cline{2-5}
 & \multirow{5}{*}{Discr. level 0.95} & $\text{FDP}$ & 0.284 (0.002) & 0.157 (0.001) \\
 & & $\mathbf{\widehat{\textbf{FDP}}}$ bias & 0.018 (0.008) & 0.033 (0.003) \\
 & & $\text{prob}\widehat{\text{FDP}}$ bias & -0.246 (0.005) & -0.140 (0.001) \\
 & & condition \ref{synth ass eq} & $e^{-08}$ ($e^{-08}$) & $e^{-07}$ ($e^{-08}$) \\
 & & AUC synth & 0.528 (0.003) & 0.530 (0.004) \\
\hline
\end{tabular}
\caption{Bias of $\hatFDP$ and of $\text{prob}\widehat{\text{FDP}}$, true FDP and order of magnitude of the difference between the proportions of $FP_{\text{synth}}$ and $FP$ ((`condition \ref{synth ass eq}')) to assess \cref{synth ass eq}; mean and standard error or standard deviation over 10 procedures. Simulated data with 200,000 and 100,000 records, and five partially identifying variables. We generated a synthetic set equal to 10\% of the largest set.}
\label{tab_scalability}
\end{table}

In \Cref{tab_scalability}, we simulated data with five discretised linkage variables with different distributions and create links at random. File \smash{$\mathcal{A}$} and file \smash{$\mathcal{B}$} contain respectively 100,000 and 200,000 records, which is a setting on which \textit{SPLink} can be used on a standard computer. We compare scenarios with different overlap degrees (defined as a fraction of the smallest file \smash{$\mathcal{A}$}) and, different degrees of difficulty defined by the discrimination level of the information contained in records (proportion of unique sets of linkage variables). We synthesise data with \textit{arf}, adapted for large scales.

We present the bias of the aggregated FDP estimate (over several runs of the procedure) as well as the true value of the FDP and the probabilistic FDP estimate from \cref{probFDP} which we can derive from the output of \textit{SPLink}. We also indicate with `condition \ref{synth ass eq}' the order of magnitude of the difference between the proportions of \smash{$FP_{\text{synth}}$} and \smash{$FP$} to assess the condition from \cref{synth ass eq}. In addition, as mentioned in \Cref{sec 2 sub 3 synthesisers}, we train a classifier to distinguish real data from synthetic data. Thus `AUC synth' assessing the synthetic data quality should be around 0.5 for the synthetic data to be indistinguishable from the real data.

As expected the true FDP decreases with the increasing overlap. Even though the bias of our estimator is larger for smaller FDP, the percentage bias does not exceed 20\%. The probabilistic FDP is generally very biased and significantly underestimate the FDP. With these simulations we demonstrate the feasibility of our procedure at large scale. In all the cases presented on \Cref{tab_scalability}, our procedure yields a reliable estimate of the FDP.

\section{Robustness to assumptions}\label{sec 6 robustness}

In this section we show the robustness of our method to deviation from our key assumptions using \textit{SPLink}, which was built for the entity resolution task and thus tackle deduplication. The main assumption is that we are able to sample from the distribution of records that do not form a link. In order to do so, we assumed so far that links happen at random. \Cref{tab_robustness} shows the robustness of our method to deviation from this assumption. Moreover, we examine the robustness of our method when duplicated data are present in the data in \Cref{tab_robustness_dedup}. We use the same simulation scheme as of \Cref{sec 5 scalability}.

It is known that when there is a systematic difference between links and non-links, RL methods have higher error rates\cite{reflexRL_2020}. Therefore, as we deviate from the assumption that links happen at random, we expect the true FDP to increase. Moreover, our estimation procedure might be hampered in the process of sampling from the distribution of the non-links. Consequently, more synthetic pairs may get linked, which would introduce positive bias as \cref{synth ass eq} is violated; we expect to overestimate the FDP. Our method would likely remain useful in that case, as the estimated FDP would represent an upper bound on the true FDP. However, two other effects interplay. First, the larger \smash{$\mathcal{B}$} is (the smaller the overlap), the fewer are records from \smash{$\mathcal{B}$} that form a link, meaning that their impact on the estimated distribution becomes limited. Second, when the RL task becomes harder due to a lower discriminative power of linkage variables, their distribution narrows, making the links and non-links distributions indistinguishable\cite{theoryRL_2018}. We are thus supposedly able to sample under the distribution of non-links unless there is a relatively high discriminative power of the linkage variables and a relatively high overlap.

\begin{table}[ht]
\centering
\begin{tabular}{|c|c|c|c|c|c|c|}
\hline
\multicolumn{3}{|c|}{ } & \multicolumn{2}{c|}{links at random} & \multicolumn{2}{c|}{links depend on variables} \\
\cline{4-7}
\multicolumn{3}{|c|}{ } & overlap 0.35 & overlap 0.75 & overlap 0.35 & overlap 0.75 \\
\hline 
\multirow{12}{*}{\rotatebox{90}{$\nA << \nB$}} & \multirow{6}{*}{Discr. level 0.85} & $\text{FDP}$ & 0.655 (0.009) & 0.475 (0.009) & 0.649 (0.023) & 0.295 (0.030) \\
 & & $\mathbf{\widehat{\textbf{FDP}}}$ bias & -0.005 (0.066) & 0.046 (0.049) & -0.068 (0.07) & -0.023 (0.047) \\
 & & $\text{prob}\widehat{\text{FDP}}$ bias & -0.019 (0.013) & 0.133 (0.009) & 0.083 (0.025) & 0.091 (0.022) \\
 & & condition \ref{synth ass eq} & $e^{-06}$ ($e^{-05}$) & $e^{-05}$ ($e^{-05}$) & $e^{-05}$ ($e^{-05}$) & $e^{-06}$ ($e^{-05}$) \\
 & & AUC synth & 0.490 (0.019) & 0.499 (0.018) & 0.476 (0.024) & 0.435 (0.023) \\
 & & AUC link & 0.497 (0.016) & 0.500 (0.010) & 1 (0) & 1 (0) \\
\cline{2-7}
 & \multirow{6}{*}{Discr. level 0.95} & $\text{FDP}$ & 0.306 (0.014) & 0.170 (0.011) & 0.227 (0.018) & 0.082 (0.010) \\
 & & $\mathbf{\widehat{\textbf{FDP}}}$ bias & -0.014 (0.054) & 0.018 (0.027) & -0.053 (0.045) & -0.008 (0.022) \\
 & & $\text{prob}\widehat{\text{FDP}}$ bias & -0.028 (0.014) & 0.086 (0.012) & 0.005 (0.024) & -0.007 (0.009) \\
 & & condition \ref{synth ass eq} & $e^{-06}$ ($e^{-05}$) & $e^{-06}$ ($e^{-06}$) & $e^{-06}$ ($e^{-06}$) & $e^{-06}$ ($e^{-06}$) \\
 & & AUC synth & 0.510 (0.021) & 0.506 (0.023) & 0.492 (0.023) & 0.461 (0.024) \\
 & & AUC link & 0.498 (0.015) & 0.502 (0.013) & 1 (0) & 1 (0) \\
\hline
\multirow{12}{*}{\rotatebox{90}{$\nA = 0.90 \times \nB$}} & \multirow{6}{*}{Discr. level 0.85} & $\text{FDP}$ & 0.657 (0.006) & 0.471 (0.008) & 0.629 (0.011) & 0.351 (0.010) \\
 & & $\mathbf{\widehat{\textbf{FDP}}}$ bias & -0.015 (0.048) & 0.042 (0.042) & -0.059 (0.053) & 0.030 (0.038) \\
 & & $\text{prob}\widehat{\text{FDP}}$ bias & -0.021 (0.006) & 0.132 (0.008) & 0.077 (0.019) & 0.165 (0.022) \\
 & & condition \ref{synth ass eq} & $e^{-06}$ ($e^{-05}$) & $e^{-05}$ ($e^{-05}$) & $e^{-05}$ ($e^{-05}$) & $e^{-05}$ ($e^{-05}$) \\
 & & AUC synth & 0.502 (0.021) & 0.491 (0.016) & 0.506 (0.019) & 0.510 (0.023) \\
 & & AUC link & 0.500 (0.012) & 0.504 (0.012) & 1 (0) & 1 (0) \\
\cline{2-7}
 & \multirow{6}{*}{Discr. level 0.95} & $\text{FDP}$ & 0.305 (0.009) & 0.168 (0.007) & 0.209 (0.014) & 0.106 (0.007) \\
 & & $\mathbf{\widehat{\textbf{FDP}}}$ bias & -0.006 (0.037) & 0.019 (0.022) & -0.027 (0.039) & 0.015 (0.019) \\
 & & $\text{prob}\widehat{\text{FDP}}$ bias & -0.026 (0.011) & 0.087 (0.007) & 0.004 (0.017) & 0.028 (0.013) \\
 & & condition \ref{synth ass eq} & $e^{-06}$ ($e^{-06}$) & $e^{-06}$ ($e^{-06}$) & $e^{-06}$ ($e^{-06}$) & $e^{-06}$ ($e^{-06}$) \\
 & & AUC synth & 0.507 (0.020) & 0.510 (0.018) & 0.524 (0.021) & 0.529 (0.023) \\
 & & AUC link & 0.497 (0.012) & 0.499 (0.014) & 1 (0) & 1 (0) \\
\hline
\end{tabular}
\caption{Bias of $\hatFDP$ and of $\text{prob}\widehat{\text{FDP}}$, true FDP and order of magnitude of the difference between the proportions of $FP_{\text{synth}}$ and $FP$ (`condition \ref{synth ass eq}') to assess \cref{synth ass eq}; mean and standard error or standard deviation over 10 procedures. We generated a synthetic set equal to 10\% of the largest set. File $\mathcal{B}$ is simulated with 5,000 records and file $\mathcal{A}$ 2,000 on the top, 4,500 on the bottom. We use 5 partially identifying variables.}
\label{tab_robustness}
\end{table}

We observe in \Cref{tab_robustness} that our method is fairly robust to deviation from the `links happen at random' assumption. We compare scenarios where links happen at random and where the linkage status depends on the values of the linkage variables which exhibit different distributions. To further understand the mechanisms at stake we also compare scenarios where \smash{$\mathcal{A}$} and \smash{$\mathcal{B}$} are of similar scale and where \smash{$\mathcal{A}$} is significantly smaller than \smash{$\mathcal{B}$}. Moreover we consider varying overlap degrees (defined as a fraction of the smallest file \smash{$\mathcal{A}$}) and, different degrees of difficulty defined by the discrimination level of the information contained in records (proportion of unique sets of linkage variables). As in the previous section, we report the bias of the aggregated FDP estimate, the true value of the FDP and the probabilistic FDP estimate. We also indicate with `condition \ref{synth ass eq}' the order of magnitude of the difference between the proportions of \smash{$FP_{\text{synth}}$} and \smash{$FP$} to assess the condition from \cref{synth ass eq}. `AUC synth' measures how well a classifier distinguishes synthetic from real records. `AUC link' measures how well a classifier distinguishes records from \smash{$\mathcal{B}$} that form a link in the other file from records that do not.

When links happen at random, the true links form a random sample from the population linkage variables. This contrasts with situations where the linkage status or registration errors depend on certain variables. In the first case, records that form a link and those that do not carry different distributions with respect to their linkage variables. In the second case, RL would be hindered in linking these specific pairs exhibiting more mistakes or more missing values. While our results show that the proposed methodology is robust to cases where linkage depends on variables in many realistic settings, it may be valuable to detect such discrepancies in the data and appraise their impact on the RL task. This could be addressed as a separate task, using synthetic data to assess potential linkage bias\cite{linkage_bias_datasynthesis}. Specifically, one could identify associations among variables and examine how these relationships affect both the linkage process and the final output.

We note from \Cref{tab_robustness} that the true FDP depends on the overlap (as defined by the smallest data set) and does not change between the scenarios \smash{$\nA << \nB$} and \smash{$\nA = 0.90 \times \nB$}. It decreases with the increasing overlap and with the increasing discrimination level. Surprisingly, \textit{SPLink} performs better when links depend on the identifying features (the true FDP is lower), which is not generally the case with RL methods\cite{reflexRL_2020}. Our method is slightly affected by the meaningful links: the percentage bias of our estimator does not exceed 10\% when links happen at random while it goes up to 20\% when the linkage status depends on the linkage variables. \textit{SPLink} does not enforce the one-to-one assignment constraint required to establish coherent links. As a results the interpretation of the linkage scores is unclear. In the settings presented on \Cref{tab_robustness}, whenever the discrimination level of the linkage variables was low, the linkage scores of \textit{SPLink} were lower than 0.5. In these cases we defined the linked pairs based on the median score returned by the algorithm. We therefore do not comment on the probabilistic FDP estimate here.

Note that the different methodologies for RL (the Fellegi-Sunter model and the graphical entity resolution framework) make different assumptions and are likely to be differently affected by the discriminative power of linkage variables, registration errors, dependencies among variables, number of entities in the population and the overlap between data sets. Here, we conducted a sensitivity analysis to assess the impact of deviations from our assumptions with \textit{SPLink}; one can expect the impact of these deviations to be slightly different for other RL methods. For instance, methods based on the Fellegi-Sunter model like \textit{SPLink} might be more affected by the varying degrees of overlap and discriminative power of linkage variables but less by the differences in distribution between links and non-links. On another note, some recent RL supervised algorithms have been developed which can learn to distinguish links and non-links\cite{splink_2022, FastHash_method_2017, DeepMatcher_method_2018}. They are expected to perform well in scenarios where the distributions of linkage variables differ. On such algorithms, our estimation procedure may not perform effectively.

So far in this work, we assumed the data to be deduplicated (all records within a data set belong to different underlying entities). In \Cref{tab_robustness_dedup} we demonstrate the robustness of our methodology to the presence of duplicated records in the data. We use the same simulation scheme as previously and we add 5\% of duplicates in each data source (2.5\% of the duplicates form a link in the other file and 2.5\% do not).

\begin{table}[ht]
\centering
\begin{tabular}{|c|c|c|c|c|}
\hline
\multicolumn{3}{|c|}{ } & overlap 0.35 & overlap 0.75  \\
\hline 
\multirow{12}{*}{\rotatebox{90}{Discr. level 0.95}} & \multirow{5}{*}{$\nA << \nB$} & $\text{FDP}$ & 0.234 (0.013) & 0.165 (0.013)  \\
 & & $\mathbf{\widehat{\textbf{FDP}}}$ bias & -0.010 (0.052) & 0.008 (0.035)  \\
 & & $\text{prob}\widehat{\text{FDP}}$ bias & 0.030 (0.014) & 0.092 (0.016) \\
 & & condition \ref{synth ass eq} & $e^{-06}$ ($e^{-05}$) & $e^{-06}$ ($e^{-05}$) \\
 & & AUC synth & 0.507 (0.020) & 0.512 (0.022) \\
\cline{2-5}
 & \multirow{5}{*}{$\nA = 0.90 \times \nB$} & $\text{FDP}$ & 0.276 (0.013) & 0.180 (0.008) \\
 & & $\mathbf{\widehat{\textbf{FDP}}}$ bias & -0.019 (0.042) & 0.008 (0.022) \\
 & & $\text{prob}\widehat{\text{FDP}}$ bias & -0.002 (0.016) & 0.077 (0.009) \\
 & & condition \ref{synth ass eq} & $e^{-06}$ ($e^{-05}$) & $e^{-06}$ ($e^{-06}$) \\
 & & AUC synth & 0.515 (0.016) & 0.515 (0.017) \\
\hline
\end{tabular}
\caption{Bias of $\hatFDP$ and of $\text{prob}\widehat{\text{FDP}}$, true FDP and order of magnitude of the difference between the proportions of $FP_{\text{synth}}$ and $FP$ (`condition \ref{synth ass eq}') to assess \cref{synth ass eq}; mean and standard error or standard deviation over 10 procedures. File $\mathcal{B}$ is simulated with 5,000 records and file $\mathcal{A}$ 2,000 on the top, 4,500 on the bottom. We use 5 partially identifying variables. We duplicated 2.5\% of records that form a link and 2.5\% of records that do not in each data source. We generated a synthetic set equal to 10\% of the largest set.}
\label{tab_robustness_dedup}
\end{table}

We observe in \Cref{tab_robustness_dedup} that our method is robust to the presence of duplicated data. When comparing results to \Cref{tab_robustness} for the same settings we can see that \textit{SPLink} becomes more conservative and therefore has a lower true FDP. We do not notice strong differences when the overlap degree is high, however the bias of our procedure is larger when the overlap degree is low. Nevertheless our estimation of the FDP is reliable in all cases. Like in \Cref{tab_robustness} the percentage bias of our procedure remains below 10\%.

\section{Applications}\label{sec 7 applications}

We evaluate our estimation method on Italian census data (SHIW) as well as on American data from a longitudinal survey (NLTCS), which provide a unique identifier. We then apply the method on Dutch perinatal data (PRN). Our procedure is generic, and can be applied to any RL algorithm, facilitating the comparison of estimation performance among RL methods. We thus show the performance of the aggregated FDP estimate (average over several run of our approach for false discovery estimation in RL) for different RL methods, data synthesisers and data sets in \Cref{sec 7 sub 1 estimation}. Then, we expand on the improvement of inference results enabled by optimising the RL algorithm thanks to our FDP estimation procedure in \Cref{sec 7 sub 2 inference}.

\subsection{FDP estimation on real data applications} \label{sec 7 sub 1 estimation}

The SHIW data come from a survey that has been conducted every two year since 1989 on the Italian population. The data are made available by the Bank of Italy for research purposes\cite{SHIW_BancadItalia_1989}. We link the data sets of 2016 and 2020, with approximately 15,000 and 16,500 records respectively, of which 6,500 are common to both files. A unique identifier can be inferred from the family and member identifiers in each sample, so that we can deduce the true linkage structure. We use the sex, birth year, regional code, marital status and education level to link the records. A simplistic matching of the records that carry identical information in all these linkage variables results in 18,000 linked pairs (among them 14,000 are false positive and 4,000 are true positive) which correspond to a FDP of about 0.77. With this matching, all false positives are due to matching identifying information between records that pertain to distinct individuals, witnessing the difficulty of the task.

The NLTCS data come from a longitudinal study on the elderly population health in the United States. The survey was sponsored by the National Institute of Aging and was conducted by the Duke University Center for Demographic Studies. They are available under request to the National Archive of Computerized Data on Aging (NACDA). We link the data sets of 1982 and 1994, with approximately 20,500 and 9,500 records respectively, of which 7,500 are common to both files. We use the birth month and year, the sex, and the state code to link the records; a unique identifier is provided, useful to evaluate our procedure. 
A simplistic matching, which links records with identical values across all linkage variables, leads to 20,000 linked pairs (of which 12,000 are false positive and 7,000 are true positive) yielding a FDP of about 0.60, demonstrating the difficulty of the task.

\begin{figure}
    \centering
    \begin{tikzpicture}[scale=0.75]
    \pgfplotsset{
    width=0.675\textwidth,
    height=0.225\textwidth}
    \begin{axis}[
    ymin=0, ymax=1,
    xmin=0, xmax=21, 
    xtick={1,2,3,4,5,6,7,8,9,10,11,12,13,14,15,16,17,18,19,20},
    xticklabels={1,,,,5,,,,,10,,,,,15,,,,,20},
    ylabel={\textit{BRL}},
    ylabel style={sloped like y axis}
    ]
    \addplot [
            only marks,
            mark=*,
            error bars/.cd,
                y dir=both,y explicit,
        ] table [x=x,y=y,y error=err] {
            x y err
            0.85 0.2398821 0.031219915 \\
            1.85 0.3619160 0.04517292 \\
            2.85 0.1272671 0.014594802 \\
            3.85 0.2866640 0.01515519 \\
            4.85 0.1720273 0.01392547 \\
            5.85 0.2689806 0.008863381 \\
            6.85 0.2454793 0.01350323 \\
            7.85 0.3142171 0.029722745 \\
            8.85 0.3232433 0.018091936 \\
            9.85 0.2165402 0.01272334 \\
            10.85 0.2525150 0.009618083 \\
            11.85 0.1644716 0.013705298 \\
            12.85 0.3734441 0.01993481 \\
            13.85 0.1805584 0.01334371 \\
            14.85 0.3085682 0.021238365 \\
            15.85 0.3428725 0.01502748 \\
            16.85 0.2559431 0.02594376 \\
            17.85 0.2459132 0.03991464 \\
            18.85 0.3125079 0.058693484 \\
            19.85 0.2486962 0.014468829 \\
        };
    \addplot [
            only marks,
            mark=*,
            mark options={draw=black,fill=white},
            error bars/.cd,
                y dir=both,y explicit,
        ] table [x=x,y=y,y error=err] {
            x y err
            1.15 0.1563787 0.04240706 \\
            2.15 0.3427208 0.05994513 \\
            3.15 0.1520569 0.04108686 \\
            4.15 0.2399441 0.04085866 \\
            5.15 0.3078326 0.06810429 \\
            6.15 0.2619957 0.03713059 \\
            7.15 0.2125713 0.02778517 \\
            8.15 0.1466600 0.03911307 \\
            9.15 0.2626522 0.03236323 \\
            10.15 0.1781300 0.02912712 \\
            11.15 0.1282438 0.02359989 \\
            12.15 0.2578038 0.04995235 \\
            13.15 0.2531868 0.02777924 \\
            14.15 0.1867781 0.03283747 \\
            15.15 0.3043238 0.03197774 \\
            16.15 0.3017721 0.05021948 \\
            17.15 0.2232301 0.04383785 \\
            18.15 0.3840288 0.04523807 \\
            19.15 0.2853621 0.07727791 \\
            20.15 0.2495433 0.03371301 \\
        };
    \end{axis}
    \end{tikzpicture}
    \begin{tikzpicture}[scale=0.75]
    \pgfplotsset{
    width=0.675\textwidth,
    height=0.225\textwidth}
    \begin{axis}[
    ymin=0, ymax=1,
    xmin=0, xmax=13, 
    xtick={1,2,3,4,5,6,7,8,9,10,11,12},
    xticklabels={,,3,,,6,,,9,,,12},
    legend entries={FDP, $\widehat{\text{FDP}}$},
    legend style={draw=none, at={(0.95,0.9)}},
    ylabel style={sloped like y axis}
    ]
    \addplot [
            only marks,
            mark=*,
            error bars/.cd,
                y dir=both,y explicit,
        ] table [x=x,y=y,y error=err] {
            x y err
            0.85 0.07023315 0.004739768 \\
            5.85 0.07095299 0.005392544 \\
            6.85 0.08897847 0.005299858 \\
            7.85 0.08454954 0.005478724 \\
            10.85 0.06713134 0.006567434 \\
        };
    \addplot [
            only marks,
            mark=*,
            mark options={draw=black,fill=white},
            error bars/.cd,
                y dir=both,y explicit,
        ] table [x=x,y=y,y error=err] {
            x y err
            1.15 0.02547824 0.006476915 \\
            2.15 0 0 \\
            3.15 0 0 \\
            4.15 0 0 \\
            5.15 0 0 \\
            6.15 0.02940501 0.006779911 \\
            7.15 0.02534251 0.007735606 \\
            8.15 0.01835351 0.006257607 \\
            9.15 0 0 \\
            10.15 0 0 \\
            11.15 0.03992635 0.009826422 \\
            12.15 0 0 \\
        };
    \end{axis}
    \end{tikzpicture}
    \\
    \begin{tikzpicture}[scale=0.75]
    \pgfplotsset{
    width=0.675\textwidth,
    height=0.225\textwidth}
    \begin{axis}[
    ymin=0, ymax=1,
    xmin=0, xmax=21, 
    xtick={1,2,3,4,5,6,7,8,9,10,11,12,13,14,15,16,17,18,19,20},
    xticklabels={1,,,,5,,,,,10,,,,,15,,,,,20},
    ylabel={\textit{FastLink}},
    ylabel style={sloped like y axis}
    ]
    \addplot [
            only marks,
            mark=*,
            error bars/.cd,
                y dir=both,y explicit,
        ] table [x=x,y=y,y error=err] {
            x y err
            0.85 0.7254631 0.007481903 \\
            1.85 0.3075145 0.03011186 \\
            2.85 0.7336064 0.004674143 \\
            3.85 0.6327448 0.01229959 \\
            4.85 0.7399252 0.01262291 \\
            5.85 0.5256321 0.010319363 \\
            6.85 0.5954387 0.01590273 \\
            7.85 0.7402334 0.005433166 \\
            8.85 0.6818398 0.009308492 \\
            9.85 0.5645193 0.01109386 \\
            10.85 0.5307887 0.014356029 \\
            11.85 0.6768628 0.008904083 \\
            12.85 0.6220515 0.01193059 \\
            13.85 0.3637592 0.01301327 \\
            14.85 0.7534041 0.006429471 \\
            15.85 0.7563525 0.00668746 \\
            16.85 0.4353334 0.01479490 \\
            17.85 0.5831254 0.01432483 \\
            18.85 0.8291045 0.007134661 \\
            19.85 0.6302461 0.007910649 \\
        };
    \addplot [
            only marks,
            mark=*,
            mark options={draw=black,fill=white},
            error bars/.cd,
                y dir=both,y explicit,
        ] table [x=x,y=y,y error=err] {
            x y err
            1.15 0.7285128 0.01715747 \\
            2.15 0.5283476 0.06343063 \\
            3.15 0.7960356 0.02047941 \\
            4.15 0.7021164 0.03149262 \\
            5.15 0.5645309 0.02077961 \\
            6.15 0.4684735 0.02998771 \\
            7.15 0.6288298 0.03389640 \\
            8.15 0.6430229 0.01873213 \\
            9.15 0.7280060 0.01967889 \\
            10.15 0.7078837 0.02056378 \\
            11.15 0.5548796 0.02415131 \\
            12.15 0.7274027 0.02415714 \\
            13.15 0.6577822 0.02627270 \\
            14.15 0.4863690 0.03323780 \\
            15.15 0.7953270 0.02161705 \\
            16.15 0.6555256 0.02306299 \\
            17.15 0.4989380 0.03529179 \\
            18.15 0.5825334 0.02586023 \\
            19.15 0.7677608 0.01828562 \\
            20.15 0.7358623 0.02034018 \\
        };
    \end{axis}
    \end{tikzpicture}
    \begin{tikzpicture}[scale=0.75]
    \pgfplotsset{
    width=0.675\textwidth,
    height=0.225\textwidth}
    \begin{axis}[
    ymin=0, ymax=1,
    xmin=0, xmax=13, 
    xtick={1,2,3,4,5,6,7,8,9,10,11,12},
    xticklabels={,,3,,,6,,,9,,,12},
    ylabel style={sloped like y axis}
    ]
    \addplot [
            only marks,
            mark=*,
            error bars/.cd,
                y dir=both,y explicit,
        ] table [x=x,y=y,y error=err] {
            x y err
            0.85 0.39337765 0.011895013 \\
            1.85 0.5843230 0.01160034 \\
            2.85 0.51974114 0.008267969 \\
            3.85 0.5657096 0.009762709 \\
            4.85 0.5426449 0.01865461 \\
            5.85 0.44473087 0.011259394 \\
            6.85 0.39819326 0.006276712 \\
            7.85 0.36381443 0.010683763 \\
            8.85 0.5342362 0.01007447 \\
            9.85 0.6038587 0.013057840 \\
            10.85 0.21172963 0.017423561 \\
            11.85 0.6768628 0.018233927 \\
        };
    \addplot [
            only marks,
            mark=*,
            mark options={draw=black,fill=white},
            error bars/.cd,
                y dir=both,y explicit,
        ] table [x=x,y=y,y error=err] {
            x y err
            1.15 0.59558885 0.020544249 \\
            2.15 0.5860163 0.02298912 \\
            3.15 0.7386120 0.01963728 \\
            4.15 0.6107566 0.01682606 \\
            5.15 0.71008716 0.02359764 \\
            6.15 0.56243130 0.013555420 \\
            7.15 0.59949632 0.027986919 \\
            8.15 0.48666499 0.013527819 \\
            9.15 0.6654341 0.01927862 \\
            10.15 0.77922001 0.021927477 \\
            11.15 0.43553882 0.027551835 \\
            12.15 0.71072381 0.02388566 \\
        };
    \end{axis}
    \end{tikzpicture}
    \\
    \begin{tikzpicture}[scale=0.75]
    \pgfplotsset{
    width=0.675\textwidth,
    height=0.225\textwidth}
    \begin{axis}[
    ymin=0, ymax=1,
    xmin=0, xmax=21, 
    xtick={1,2,3,4,5,6,7,8,9,10,11,12,13,14,15,16,17,18,19,20},
    xticklabels={1,,,,5,,,,,10,,,,,15,,,,,20},
    xlabel={SHIW regions},
    ylabel={\textit{FlexRL}},
    ylabel style={sloped like y axis}
    ]
    \addplot [
            only marks,
            mark=*,
            error bars/.cd,
                y dir=both,y explicit,
        ] table [x=x,y=y,y error=err] {
            x y err
            0.85 0.3695858 0.023937748 \\
            1.85 0.1309316 0.03793775 \\
            2.85 0.2764851 0.016397342 \\
            3.85 0.3449719 0.02693756 \\
            4.85 0.3484726 0.01495740 \\
            5.85 0.2972034 0.024115219 \\
            6.85 0.3056726 0.01556247 \\
            7.85 0.3290385 0.025188613 \\
            8.85 0.3877031 0.017222493 \\
            9.85 0.3519442 0.01778825 \\
            10.85 0.3014713 0.020905274 \\
            11.85 0.2671404 0.016679016 \\
            12.85 0.3613453 0.01783770 \\
            13.85 0.2354507 0.03658247 \\
            14.85 0.4085356 0.021850812 \\
            15.85 0.4386308 0.02763137 \\
            16.85 0.3067033 0.03214390 \\
            17.85 0.3439004 0.02170628 \\
            18.85 0.5524608 0.017940955 \\
            19.85 0.3316726 0.020218438 \\
        };
    \addplot [
            only marks,
            mark=*,
            mark options={draw=black,fill=white},
            error bars/.cd,
                y dir=both,y explicit,
        ] table [x=x,y=y,y error=err] {
            x y err
            1.15 0.3563579 0.02617215 \\
            2.15 0.2547152 0.06243450 \\
            3.15 0.3289417 0.02902680 \\
            4.15 0.3488484 0.05090620 \\
            5.15 0.4697063 0.03981120 \\
            6.15 0.3425391 0.03382157 \\
            7.15 0.2917482 0.03274473 \\
            8.15 0.2659462 0.02773323 \\
            9.15 0.3169208 0.02533640 \\
            10.15 0.2941540 0.03604890 \\
            11.15 0.2016668 0.02394079 \\
            12.15 0.3642880 0.02954355 \\
            13.15 0.2726268 0.03856629 \\
            14.15 0.2469008 0.04155640 \\
            15.15 0.3716396 0.02989999 \\
            16.15 0.4041780 0.04681308 \\
            17.15 0.2091408 0.03816260 \\
            18.15 0.4152517 0.04339346 \\
            19.15 0.3957034 0.04675033 \\
            20.15 0.2912104 0.02666180 \\
        };
    \end{axis}
    \end{tikzpicture}
    \begin{tikzpicture}[scale=0.75]
    \pgfplotsset{
    width=0.675\textwidth,
    height=0.225\textwidth}
    \begin{axis}[
    ymin=0, ymax=1,
    xmin=0, xmax=13, 
    xtick={1,2,3,4,5,6,7,8,9,10,11,12},
    xticklabels={,,3,,,6,,,9,,,12},
    xlabel={NLTCS census regional offices},
    ylabel style={sloped like y axis}
    ]
    \addplot [
            only marks,
            mark=*,
            error bars/.cd,
                y dir=both,y explicit,
        ] table [x=x,y=y,y error=err] {
            x y err
            0.85 0.08519969 0.010256770 \\
            1.85 0.1631551 0.03527969 \\
            2.85 0.08051297 0.010310557 \\
            3.85 0.1959409 0.023342651 \\
            4.85 0.1761428 0.02269462 \\
            5.85 0.08224646 0.012738409 \\
            6.85 0.09789314 0.007588523 \\
            7.85 0.09182094 0.010405507 \\
            8.85 0.1295111 0.01542098 \\
            9.85 0.1020756 0.009358652 \\
            10.85 0.10354945 0.015689522 \\
            11.85 0.002380952 0.007328417 \\
        };
    \addplot [
            only marks,
            mark=*,
            mark options={draw=black,fill=white},
            error bars/.cd,
                y dir=both,y explicit,
        ] table [x=x,y=y,y error=err] {
            x y err
            1.15 0.05114240 0.011023715 \\
            2.15 0.1246617 0.04413078 \\
            3.15 0.1197335 0.02003286 \\
            4.15 0.1461385 0.06719707 \\
            5.15 0.08553875 0.02319617 \\
            6.15 0.06932584 0.009087154 \\
            7.15 0.05617658 0.011058809 \\
            8.15 0.06691250 0.011053595 \\
            9.15 0.1230268 0.02941638 \\
            10.15 0.04238512 0.009485337 \\
            11.15 0.14227824 0.014307377 \\
            12.15 0.08957478 0.03391834 \\
        };
    \end{axis}
    \end{tikzpicture}
    \caption{$\hatFDP$ (mean and standard error) and true FDP (mean and standard deviation). We truncated the few estimates exceeding 1. We generate for each region a synthetic set of 10\% of the biggest set. \textbf{SHIW (regions)} with between 60 and 1791 records, on the left. The overlapping set, common to both data sources, is between 34\% and 64\% of the smallest set. Only 60 and 62 records come from region 2. Other regions gather at least 150 records in both sources. \textbf{NLTCS (census regional offices)} with between 502 and 2,860 records, on the right. The overlapping set, common to both data sources, is between 67\% and 77\% of the smallest set.}
    \label{fig-SHIW-NLTCS-Regions}
\end{figure}
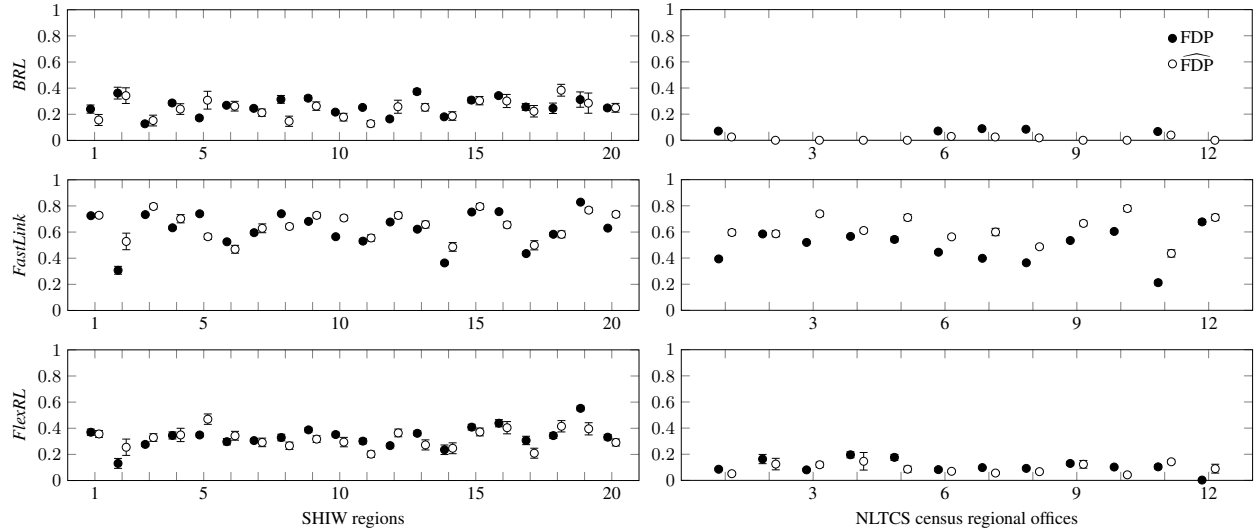

First, we estimate the FDP for three RL methods and two data synthesisers on the areas subsets of the SHIW data (blocking on North, Centre, South of Italy) in \Cref{tab-SHIW-Areas}. As we observe that \textit{synthpop} and \textit{arf} perform similarly well (the AUC synth assessing the synthetic data quality for both methods are similar), we then show estimation results with \textit{synthpop} on the full SHIW and NLTCS data in \Cref{tab-SHIW-NLTCS-Full}, as well as on the regional subsets of the SHIW and NLTCS data (blocking on 20 regions of Italy and on 12 census regional offices of the U.S.) in \Cref{fig-SHIW-NLTCS-Regions}. 

We repeat the procedure for false discovery estimation in RL ten times. We provide in the following tables and figures (\Cref{tab-SHIW-Areas}, \Cref{tab-SHIW-NLTCS-Full}, \Cref{fig-SHIW-NLTCS-Regions}) the mean and standard error over the sample of FDP estimates obtained.


\begin{table}[!ht]
    \centering
    \begin{tabular}{|c||c|c||c|c||c|c|}
      \hline
      \multicolumn{1}{|c||}{\textit{BRL}} & \textit{synthpop} & \textit{arf} & \textit{synthpop} & \textit{arf} & \textit{synthpop} & \textit{arf} \\
      \hline
       $\text{FDP}$ & 0.073 (0.021) & 0.076 (0.028) & 0.222 (0.016) & 0.218 (0.017) & 0.188 (0.11) & 0.187 (0.012) \\
       $\mathbf{\widehat{\textbf{FDP}}}$ bias & 0.093 (0.031) & 0.019 (0.070) & -0.001 (0.012) & -0.049 (0.008) & -0.001 (0.024) & -0.077 (0.022) \\
       $\text{prob}\widehat{\text{FDP}}$ bias & Non Available & Non Available & Non Available & Non Available & Non Available & Non Available \\
       condition \ref{synth ass eq} & $e^{-07}$ ($e^{-07}$) & $e^{-07}$ ($e^{-07}$) & $e^{-06}$ ($e^{-07}$) & $e^{-06}$ ($e^{-06}$) & $e^{-07}$ ($e^{-07}$) & $e^{-07}$ ($e^{-07}$) \\
      \hline
      \hline
      \multicolumn{1}{|c||}{\textit{FastLink}} & \textit{synthpop} & \textit{arf} & \textit{synthpop} & \textit{arf} & \textit{synthpop} & \textit{arf}\\
      \hline
       $\text{FDP}$ & 0.791 (0.003) & 0.789 (0.004) & 0.727 (0.006) & 0.727 (0.004) & 0.741 (0.004) & 0.745 (0.005) \\
       $\mathbf{\widehat{\textbf{FDP}}}$ bias & -0.043 (0.012) & -0.009 (0.012) & 0.002 (0.008) & -0.014 (0.004) & 0.011 (0.008) & 0.010 (0.011) \\
       $\text{prob}\widehat{\text{FDP}}$ bias & -0.591 (0.001) & -0.589 (0.001) & -0.494 (0.001) & -0.495 (0.000) & -0.555 (0.001) & -0.555 (0.005) \\
       condition \ref{synth ass eq} & $e^{-05}$ ($e^{-06}$) & $e^{-06}$ ($e^{-06}$) & $e^{-05}$ ($e^{-05}$) & $e^{-06}$ ($e^{-06}$) & $e^{-06}$ ($e^{-06}$) & $e^{-06}$ ($e^{-06}$) \\
      \hline
      \hline
      \multicolumn{1}{|c||}{\textit{FlexRL}} & \textit{synthpop} & \textit{arf} & \textit{synthpop} & \textit{arf} & \textit{synthpop} & \textit{arf} \\
      \hline
       $\text{FDP}$ & 0.386 (0.019) & 0.371 (0.013) & 0.405 (0.017) & 0.413 (0.019) & 0.410 (0.014) & 0.402 (0.016) \\
       $\mathbf{\widehat{\textbf{FDP}}}$ bias & 0.009 (0.019) & 0.025 (0.031) & -0.023 (0.018) & -0.077 (0.007) & -0.028 (0.016) & -0.052 (0.029) \\
       $\text{prob}\widehat{\text{FDP}}$ bias & -0.052 (0.005) & -0.031 (0.003) & -0.082 (0.002) & -0.086 (0.002) & -0.098 (0.002) & -0.090 (0.016) \\
       condition \ref{synth ass eq} & $e^{-06}$ ($e^{-07}$) & $e^{-07}$ ($e^{-07}$) & $e^{-06}$ ($e^{-06}$) & $e^{-06}$ ($e^{-06}$) & $e^{-06}$ ($e^{-06}$) & $e^{-06}$ ($e^{-06}$) \\
      \hline
      \hline
      \multicolumn{1}{|c||}{} & \textit{synthpop} & \textit{arf} & \textit{synthpop} & \textit{arf} & \textit{synthpop} & \textit{arf} \\
      \hline
      AUC synth & \multicolumn{1}{c}{0.502 (0.025)} & \multicolumn{1}{c||}{0.503 (0.019)} & \multicolumn{1}{c}{0.488 (0.034)} & \multicolumn{1}{c||}{0.478 (0.026)} & \multicolumn{1}{c}{0.495 (0.010)} & \multicolumn{1}{c|}{0.515 (0.024)} \\
      AUC link & \multicolumn{2}{c||}{0.631 (0.007)} & \multicolumn{2}{c||}{0.597 (0.011) } & \multicolumn{2}{c|}{0.612 (0.008) } \\
    \hline
    \end{tabular}
    \caption{Bias of $\hatFDP$ and of $\text{prob}\widehat{\text{FDP}}$, true FDP and order of magnitude of the difference between the proportions of $FP_{\text{synth}}$ and $FP$ (`condition \ref{synth ass eq}') to assess \cref{synth ass eq}; mean and standard error or standard deviation over 10 procedures. \textbf{SHIW (North / Centre / South)} with 6,723 and 6,518 / 3,450 and 3,046 / 6,223 and 5,353 records. We generate for each area a synthetic set of 10\% of the biggest set. The overlapping set, common to both data sources, contains 2,506 / 1,385 / 2,531 records i.e.\ 38\% / 45\% / 47\% of the smallest set.}
    \label{tab-SHIW-Areas}
\end{table}


We present on \Cref{tab-SHIW-Areas} the bias of the aggregated FDP estimate as well as the true value of the FDP and, when possible the probabilistic FDP estimate from \cref{probFDP}, for the North, Centre and South subsets of the SHIW data sets. We also indicate the order of magnitude of the difference between the proportions of \smash{$FP_{\text{synth}}$} and \smash{$FP$} to assess the condition from \cref{synth ass eq}.

We notice one bad estimation of the FDP for \textit{BRL} with \textit{synthpop} on the first area (0.09 bias for a true FDP of 0.07). In all the other cases our procedure leads to a reliable estimate of the FDP. The probabilistic FDP---which may be computed on the output of \textit{FastLink} and \textit{FlexRL}---always has a larger bias and is particularly unreliable for \textit{FastLink}. The estimations obtained from synthetic data generated with \textit{arf} and \textit{synthpop} are similarly good.


\begin{table}[!ht]
    \centering
    \begin{tabular}{c}
    \parbox[t]{2mm}{\rotatebox[origin=c]{90}{\hphantom{..........}SHIW\hphantom{.........}}} \\
    \\
    \parbox[t]{2mm}{\rotatebox[origin=c]{90}{\hphantom{...}NLTCS\hphantom{..}}} \\
    \end{tabular}
    \begin{tabular}{|c|c|}
      \hline
       \multicolumn{1}{|c|}{\textit{BRL}} & \textit{synthpop} \\
      \hline
       $\text{FDP}$ &  0.23 (0.006)  \\
       $\mathbf{\widehat{\textbf{FDP}}}$ bias &  -0.035 (0.001)  \\
       $\text{prob}\widehat{\text{FDP}}$ bias & Non Available \\
       condition \ref{synth ass eq} & $e^{-07}$ ($e^{-07}$) \\
      \hline
      \hline
       \multicolumn{1}{|c|}{\textit{BRL}} & \textit{synthpop} \\
      \hline
       $\text{FDP}$ & \multirow{4}{*}{Non Available} \\
       $\mathbf{\widehat{\textbf{FDP}}}$ bias & \hphantom{-0.035 (0.001)}  \\
       $\text{prob}\widehat{\text{FDP}}$ bias & \\
       condition \ref{synth ass eq} &  \\
      \hline
    \end{tabular}
        \begin{tabular}{|c|c|}
      \hline
      \multicolumn{1}{|c|}{\textit{FastLink}} & \textit{synthpop} \\
      \hline
       $\text{FDP}$ &  0.70 (0.003)  \\
       $\mathbf{\widehat{\textbf{FDP}}}$ bias & -0.062 (0.002) \\
       $\text{prob}\widehat{\text{FDP}}$ bias & -0.57 (0.001) \\
       condition \ref{synth ass eq} & $e^{-06}$ ($e^{-06}$) \\
      \hline
      \hline
      \multicolumn{1}{|c|}{\textit{FastLink}} & \textit{synthpop} \\
      \hline
       $\text{FDP}$ & 0.494 (0.005)  \\
       $\mathbf{\widehat{\textbf{FDP}}}$ bias & 0.139 (0.008) \\
       $\text{prob}\widehat{\text{FDP}}$ bias & -0.145 (0.005) \\
       condition \ref{synth ass eq} & $e^{-06}$ ($e^{-06}$) \\
      \hline
    \end{tabular}
    \begin{tabular}{|c|c|}
      \hline
      \multicolumn{1}{|c|}{\textit{FlexRL}} & \textit{synthpop} \\
      \hline
       $\text{FDP}$ &  0.42 (0.001) \\
       $\mathbf{\widehat{\textbf{FDP}}}$ bias & \hphantom{-}0.009 (0.008) \\
       $\text{prob}\widehat{\text{FDP}}$ bias & -0.10 (0.003) \\
       condition \ref{synth ass eq} & $e^{-05}$ ($e^{-06}$)  \\
      \hline
      \hline
      \multicolumn{1}{|c|}{\textit{FlexRL}} & \textit{synthpop} \\
      \hline
       $\text{FDP}$ &  0.10 (0.004) \\
       $\mathbf{\widehat{\textbf{FDP}}}$ bias & -0.019 (0.007) \\
       $\text{prob}\widehat{\text{FDP}}$ bias & 0.189 (0.076) \\
       condition \ref{synth ass eq} & $e^{-05}$ ($e^{-06}$)  \\
      \hline
    \end{tabular}
    \caption{Bias of $\hatFDP$ and of $\text{prob}\widehat{\text{FDP}}$, true FDP and order of magnitude of the difference between the proportions of $FP_{\text{synth}}$ and $FP$ (`condition \ref{synth ass eq}') to assess \cref{synth ass eq}; mean and standard error or standard deviation over 10 procedures. We generated a synthetic set equal to 10\% of the largest set. \textbf{SHIW (full data)} (top) from 2016 and 2020 containing respectively 16,445 and 14,917 records. The overlapping set, common to both data sources, contains 6,430 records i.e.\ 43\% of the smallest set. AUC synth: 0.490 (0.008) and AUC link: 0.613 (0.003). \textbf{NLTCS (full data)} (bottom) from 1982 and 1994 containing respectively 20,484 and 9,532 records. The overlapping set, common to both data sources, contains 7,612 records i.e.\ 80\% of the smallest set. AUC synth: 0.688 (0.012) and AUC link: 0.739 (0.004). \textit{BRL} did not link any record from the NLTCS data.}
    \label{tab-SHIW-NLTCS-Full}
\end{table}

We present similar results on \Cref{tab-SHIW-NLTCS-Full} for the full SHIW and NLTCS data sets. \textit{FastLink} links a lot of records compared to the other methods, explaining the high FDPs. The probabilistic FDP (computed for \textit{FastLink} and \textit{FlexRL}) always has a larger bias than our estimate. We only rely on four linkage variables to link the NLTCS data, which may explain the relatively bad FDP estimation of \textit{FastLink} (0.15 bias) and that \textit{BRL} did not link any record.


We present on \Cref{fig-SHIW-NLTCS-Regions} the true value of the FDP and the aggregated FDP estimate for each regional subset of the SHIW data sets and for each census regional office subset of the NLTCS data sets. The second region of the SHIW data contain less than a hundred observations in both sources, probably explaining the bad estimation of the FDP with \textit{FastLink} compared to the other regions. Like on the full NLTCS data, \textit{BRL} does not link any records on several of the census regional offices subsets. We truncated the few estimates exceeding 1. The FDP tend to be overestimated with \textit{FastLink} and underestimated with \textit{FlexRL}.

Overall, the percentage of bias (compared to the true FDP value) lays around 15\% on average over the different RL methods and data synthesisers. Our FDP estimation procedure performs well on large and small data sets (with and without blocking) and is able to provide an accurate order of magnitude of the FDP. The overlap degrees in these real data applications varied between 30\% and 80\% and as we already saw in the precedent \Cref{sec 5 scalability} and \Cref{sec 6 robustness} the bias of our estimator is not affected by the overlap. Our methodology is reliable under various real world settings with different scales, different degrees of overlap and of difficulty. The probabilistic estimate becomes more reliable with higher overlap between data sets and more discriminative linkage variables. When available, it can be assessed alongside with our approach but in general, our method remains the more accurate and applicable option.

\subsection{FDP estimation: a tool for improving inference on linked data} \label{sec 7 sub 2 inference}

Inference drawn from linked data suffers from inherent weaknesses due to the linkage process. We saw from the previous tables that the probabilistic FDP estimate of \cref{probFDP} which we can compute on \textit{FastLink} or \textit{FlexRL} is usually not reliable: its percentage bias (compared to the true FDP value) varies between 20\% and 200\%. This estimate is derived from the RL model, which in turn relies on simplifying assumptions. As such models are rarely well calibrated in real world settings, the resulting bias in the probabilistic estimate is to be expected. Hence the need for a procedure for false discovery estimation in RL, that is independent of the linkage process itself, in order to estimate the FDP of any linkage method in any context, and to inform the analysis on the reliability of its data. Our procedure answers these needs. We are able to evaluate the RL error depending on the RL parameters. Therefore, by tightening the parameters we can effectively tune the RL algorithm in order to minimise the FDP and, we can assess the reliability of the optimal set of linked records obtained.

As mentioned in \Cref{sec 2 sub 1 rl method}, RL methods have different ways of classifying linked and non-linked pairs. \textit{FastLink} and \textit{FlexRL} return pairs of records and their linkage scores, allowing explicit `tuning' of the threshold \smash{$\xi$}. By increasing its value one can lower the FDP (hence $\hatFDP$). \textit{BRL} does not allow explicit adjustment of the threshold $\xi$ on linkage scores but some equivalent parameters which control the set of linked records returned. For this method, the FDP (hence \smash{$\hatFDP$}) may be lowered by taking \smash{$\lambda_{\text{FNM}} < \lambda_{\text{FM1}}, \lambda_{\text{FM2}}$} while keeping \smash{$0 < \lambda_{\text{FNM}} \leq \lambda_{\text{FM1}}$} and \smash{$\lambda_{\text{FNM}} + \lambda_{\text{FM1}} \leq \lambda_{\text{FM2}}$}\cite{sadinle_bay_bipartite_RL_2017}. For instance, \smash{$\lambda_{\text{FNM}}$}, which default value is \smash{$1$} may be lowered. The default values for \textit{BRL} parameters: \smash{$\lambda_{\text{FNM}} = \lambda_{\text{FM1}} = 1, \lambda_{\text{FM1}} = 2$} are equivalent\cite{sadinle_bay_bipartite_RL_2017} to the default threshold value \smash{$\xi=0.5$} on the linkage scores for \textit{FastLink} and \textit{FlexRL}.

In this section we show results from inference done with the default parameters of the RL methods and the associated \smash{$\hatFDP$}. We compare it with the results obtained after adjusting the parameters of the RL methods in order to obtain a lower \smash{$\hatFDP$}. We target a \smash{$\hatFDP$} lower than 10\%; however in some cases, the optimal set of linked pairs does not allow to reach such threshold. In these cases, neither the linkage nor the inference is reliable. Otherwise optimising the linkage and assessing its error allows to improve the inference. It is important to note that by choosing more stringent parameters, we also loose observations, affecting the statistical power of the inference results. Therefore we present results obtained from the full data sets (with blocking when relevant) to obtain enough linked observations.


First, we apply RL to assemble the first and second born children of the PRN data and we estimate the pre-term birth risk on the linked data using characteristics of the mother and of the previous delivery. The PRN data are available upon request from the Dutch PRN organisation. Information was collected at the scope of the pregnancy. We have access to the delivery date, birth order of the baby among the siblings, sex of the baby, pregnancy duration, an indicator for congenital malformation, an indicator for twin pregnancy, an indicator for the use of Assisted Reproductive Technology (ART), as well as characteristics on the mother namely, the birth date, date of the previous delivery (if any), ethnicity, postal code, and an indicator for low socio-economic status. The data source initially contained nearly 2 million observations, we therefore selected a smaller sample with complete data, most common ethnicity, no congenital malformation, single pregnancy, mother aged between 20 and 45 years old, low socio-economic status. To link the data, we use the mother birth date and the siblings birth dates (birth date of the first born child and birth date of the previous delivery for the second born child), as well as the postal code. We present the results on \Cref{tab-PRN-res}.

\begin{table}[!ht]
    \centering
    \begin{tabular}{|c|c|}
      \hline
      \textit{BRL} & default \\
      \hline
       $\widehat{\text{FDP}}$ & 0.01 \\
       intercept & 6.64 (0.83) $\star$ \\
       int btw pregnancies & -0.00 (0.05) \\
       mother age at 1st & -0.04 (0.02) $\star$ \\
       duration pregnancy 1 & -0.21 (0.02) $\star$ \\
       ART pregnancy 1 & -0.13 (0.18) \\
      \hline
    \end{tabular}
    \begin{tabular}{|c|c|}
      \hline
      \textit{FastLink} & default \\
      \hline
       $\widehat{\text{FDP}}$ & 0.01 \\
       intercept & 6.60 (0.83) $\star$ \\
       int btw pregnancies & -0.01 (0.05) \\
       mother age at 1st & -0.04 (0.01) $\star$ \\
       duration pregnancy 1 & -0.21 (0.02) $\star$ \\
       ART pregnancy 1 & -0.14 (0.18) $\star$ \\
      \hline
    \end{tabular}
    \begin{tabular}{|c|c|}
      \hline
      \textit{FlexRL} & default \\
      \hline
       $\widehat{\text{FDP}}$ & 0.00 \\
       intercept & 6.47 (0.94) $\star$ \\
       int btw pregnancies & -0.03 (0.07) \\
       mother age at 1st & -0.04 (0.02) $\star$ \\
       duration pregnancy 1 & -0.21 (0.02) $\star$ \\
       ART pregnancy 1 & -0.32 (0.21) $\star$ \\
      \hline
    \end{tabular}
    \caption{Results of the logistic regression fitted to estimate the probability of a premature 2nd child on the PRN linked data in North-Holland. We use the mother interval between pregnancies, the mother age at the 1st delivery, the duration of the 1st delivery, the use of ART for the 1st delivery, $\star$: significantly non-zero at level 0.01.}
    \label{tab-PRN-res}
\end{table}

All the methods have a very low estimated FDP (0.01 maximum) with their default parameters, comforting the use of these linked data in the literature. They all link around 65\% of the records (as a fraction of the smallest file). We estimate the pre-term birth risk at the second delivery on the linked data using a logistic regression on the interval between pregnancies, the mother age at the first delivery, the duration of the first pregnancy and the use of ART for the first pregnancy. The coefficient of the logistic regression are similar over all the methods, except that the data linked with \textit{FlexRL} show a weaker effect of the ART during the first pregnancy on the pre-term birth risk at the second delivery.


Second, we apply RL to gather the records pertaining to the same individuals in the SHIW data. We generated an outcome in one file as a linear function of three covariates simulated in the other file for the links. Non-linked outcomes were sampled from the outcome empirical cumulative distribution function of links using inverse transform sampling. We fit a linear regression on these variables and we show how to use the FDP estimation to refine the set of linked records by potentially decreasing \smash{$\lambda_{\text{FNM}}$} in \textit{BRL} or increasing \smash{$\xi$} in \textit{FastLink} and \textit{FlexRL}. That way, we can improve the inference results. We present the results on \Cref{tab-SHIW-res}.

\begin{table}[!ht]
    \centering
    \begin{tabular}{|c|c|}
       \hline
       DGP &  \\
       \hline
       \vphantom{$\widehat{\text{FDP}}$} &  \\
       intercept & -5  \\
       $\beta_1$ & 1  \\
       $\beta_2$ & 1  \\
       $\beta_3$ & 20 \\
       \vphantom{$R^2$} & \\
      \hline
    \end{tabular}
    \begin{tabular}{|c|c|}
      \hline
      \textit{BRL} & default \\
      \hline
       $\widehat{\text{FDP}}$ & 0.09 \\
       intercept & -4.79 (0.21) $\star$ \\
       $\beta_1$ & 1.14 (0.24) $\star$ \\
       $\beta_2$ & 0.98 (0.06) $\star$ \\
       $\beta_3$ & 20.32 (0.52) $\star$ \\
       $R^2$ & 0.98 \\
      \hline
    \end{tabular}
    \begin{tabular}{|c|c|c|}
      \hline
      \textit{FastLink} & default & tuned \\
      \hline
      $\widehat{\text{FDP}}$ & 0.75 & 0.65 \\
       intercept & -4.72 (0.09) $\star$ & -4.81 (0.10) $\star$ \\
       $\beta_1$ & 0.02 (0.09) & 0.04 (0.10) \\
       $\beta_2$ & 0.07 (0.03) $\star$ & 0.10 (0.03) $\star$ \\
       $\beta_3$ & 1.84 (0.22) $\star$ & 2.27 (0.24) $\star$ \\
       $R^2$ & 0.01 & 0.01 \\
      \hline
    \end{tabular}
    \begin{tabular}{|c|c|c|}
      \hline
      \textit{FlexRL} & default & tuned \\
      \hline
      $\widehat{\text{FDP}}$ & 0.39 & 0.18 \\
       intercept & -4.84 (0.19) $\star$ & -4.62 (0.40) $\star$ \\
       $\beta_1$ & 0.76 (0.18) $\star$ & 1.12 (0.37) $\star$ \\
       $\beta_2$ & 0.64 (0.06) $\star$ & 0.80 (0.14) $\star$ \\
       $\beta_3$ & 12.24 (0.46) $\star$ & 14.99 (0.98) $\star$ \\
       $R^2$ & 0.38 & 0.58 \\
      \hline
    \end{tabular}
    \caption{Results of the linear model estimated on the SHIW linked data (with blocking on areas North, Centre, South). The left column indicates the Data Generating Process (DGP) of the bimodal outcome: $Y = -5 + 1 \cdot X_1 + 1 \cdot X_2 + 20 \cdot X_3 + \varepsilon$ with $X_1 \sim \mathcal{N}(0,1), X_2 \sim \mathcal{N}(0,3), X_3 \sim \beta(0.2,0.1)-2/3, \varepsilon \sim \mathcal{N}(0,1)$, $\star$: significantly non-zero at level 0.01.}
    \label{tab-SHIW-res}
\end{table}

The estimated FDP of \textit{BRL} of $0.09$ with its default parameters leads to a coefficient of determination of \smash{$98\%$} in the regression model; all the coefficients are well estimated. The \smash{$\hatFDP$} of \textit{FastLink} with the default threshold is \smash{$0.75$}. The strictest set of linked records we can obtained by increasing the threshold \smash{$\xi$} leads to a \smash{$\hatFDP$} of \smash{$0.65$} which is not satisfying. We can suspect here that the coefficient of determination of \smash{$1\%$} results from a signal diluted in the noise of falsely linked records. In the same fashion, \textit{FlexRL} with the default threshold leads to a \smash{$\hatFDP$} of \smash{$0.39$}, associated with a \smash{$R^2=38\%$}. By tuning this parameter we can obtain a coefficient of determination of \smash{$58\%$}, the \smash{$\hatFDP$} associated with the linkage in that case is \smash{$0.18$}.


Third, we apply RL to gather the records pertaining to the same individuals in the NLTCS data. We fit a linear model to explain the Frailty Index (FI) of individuals from 1994 using the their FI from 1982. The frailty index appears as a measure of the health and quality of life of the individuals. We define it following the methodology of previous studies\cite{Mitnitski_FI_2004, RockwoodCSHA_2004, NLTCS_FI_2006} as a proportion of deficits. Explicitly, we computed the mean over 45 indicators recording difficulty with: eating, dressing, going out, walking around, getting in/out of chairs and of bed, bathing, going to the toilets, cooking, doing easy task like laundry or washing the dishes, shopping, handling money, taking medicine, calling, reading, speaking, understanding, as well as health issues: rheumatism, paralysis, numbness, scleroses, epilepsy, cerebral palsy, Parkinson's disease, glaucoma, diabetes, cancer, constipation, trouble sleeping, headaches, overweight, arteriosclerosis, mental retard senility, history of heart attack or stroke, heart problems, hypertension, pneumonia, flu, emphysema, asthma, bronchitis, blood circulation problems, broken bones, missing fingers or toes. We present the results on \Cref{tab-NLTCS-res}.

\begin{table}[!ht]
    \centering
    \begin{tabular}{|c|c|}
      \hline
      Benchmark &  \\
      \hline
       \vphantom{$\widehat{\text{FDP}}$} &  \\
       intercept & 0.05 (0.00) $\star$ \\
       $FI82$ & 0.67 (0.02) $\star$ \\
       $R^2$ & 0.17 \\
      \hline
    \end{tabular}
    \begin{tabular}{|c|c|c|}
      \hline
      \textit{FastLink} & default & tuned \\
      \hline
       $\widehat{\text{FDP}}$ & 0.63 & 0.35 \\
       intercept & -0.05 (0.00) $\star$ & 0.05 (0.01) $\star$ \\
       $FI82$ & 0.06 & 0.51 (0.02) $\star$ \\
       $R^2$ & 0.03 & 0.11 \\
      \hline
    \end{tabular}
    \begin{tabular}{|c|c|}
      \hline
      \textit{FlexRL} & default \\
      \hline
       $\widehat{\text{FDP}}$ & 0.08 \\
       intercept & 0.05 (0.00) $\star$ \\
       $FI82$ & 0.62 (0.05) $\star$ \\
       $R^2$ & 0.14 \\
      \hline
    \end{tabular}
    \caption{Results of the linear model estimated on the NLTCS linked data. We use the Frailty Index (FI) from 1982 to predict the FI in 1994, $\star$: significantly non-zero at level 0.01. The left column shows the coefficient estimates obtained from the linear model on the true links. \textit{BRL} does not appear in the table as it does not link enough records.}
    \label{tab-NLTCS-res}
\end{table} 

The estimated FDP of \textit{FastLink} with the default threshold is \smash{$0.63$}, the linkage in that case is not reliable. By tuning this parameter we obtain a \smash{$\hatFDP$} of \smash{$0.35$} and a coefficient of determination of \smash{$11\%$} (compared to \smash{$17\%$} on the true links), the coefficients are better estimated. The \smash{$\hatFDP$} of \textit{FlexRL} of \smash{$0.08$} with the default threshold leads to a coefficient of determination of \smash{$14\%$} in the regression model, the estimated coefficients are closer to the one obtained from fitting the model on the true links.

Therefore, these examples show the importance of the ability to estimate the FDP and thereby to tune the RL parameters to obtain reliably linked data.

\section{Discussion} \label{sec 8 discussion}

Providing reliable tools for estimating the False Discovery Proportion (FDP) is essential for downstream applications of Record Linkage (RL) to be meaningfully pursued. In lack of robust method to evaluate the error when linking data through RL, the reliability of the linkage cannot be assessed. While the final decision of which FDP is acceptable for a particular analysis should rest within the researcher's hand, it is important to appraise the complexity of the RL task and to be informed on the magnitude of the error in the resulting linked data. We therefore provided an unbiased procedure to derive an estimate of the FDP in RL, which accordingly allows to optimise the RL parameters to improve the inference.

Our method proves particularly useful for complex RL tasks with weak linkage variables, where standard simplifying assumptions (i.i.d.\ records, independent comparison vectors and linking variables, no duplicates, registration errors at random) affect the linkage scores. Such context allows us to properly sample synthetic records, well representing the original records which do not form links, to delude the RL process and measure the error made when linking records. It relies on the realistic assumption that the linkage status does not depend on any observed mechanism but remains robust against deviations from it. Moreover it is not sensitive to the presence of duplicated records in the data sources. It is a generic method which can be used with any RL algorithm, hence independent of the underlying RL model. Several procedural options, which we investigated in this research work, from the synthesis of records to the choice of an estimator, have consequences on the accuracy of the method. We demonstrated robustness across a variety of scenarios and illustrated the gain of our procedure over the existing model dependent probabilistic estimate.

An adequate set of synthetic records enables a good FDP estimation. We put light on two existing methods, \textit{synthpop} and \textit{arf} whose syntheses are adapted to the FDP estimation procedure. Whereas \textit{synthpop} may be slow on large data sets with categorical data of high cardinality and complex structures, \textit{arf} takes advantage of a larger number of data. Although the performance of RL changes with the scale of the data sources, we do not recommend adapting the FDP estimation by including the contributions of synthetic records, in line with the conclusions drawn from target-decoy approaches\cite{target_decoy_fdr_2012}. The estimate we derived scales well with the different complexities coming from the RL task, it adapts to the size of the data sources, to the degree of overlap and to the discriminative power of linking variables. When the RL method provided an FDP estimate contingent on the linkage modelling, it was always highly underestimating the FDP. The procedure we proposed as an alternative for false discovery estimation is more reliable and accommodates well to real world RL settings.

Based on those results, we recommend synthesising a small set of records comprising 10\% of the number of records in the largest original source to avoid compromising the record linkage process. The number of records to synthesise should be sufficient to stabilise the estimator, as indicated by the decreasing variance, without extending so far that the variance reduction reflects a shift in the RL behaviour rather than genuine convergence of the FDP. Using the synthetic linked data to approximate the real quantity of falsely linked records allows to derive an unbiased estimate of the FDP. We argue that following this simple methodology to estimate the FDP in RL could substantially broaden the applicability of RL methods, as we have shown that it is important to tune RL to obtain a set of linked data with an acceptable FDP, which then improves inference. The ultimate optimisation of the RL parameters enabled by our procedure should balance the estimated FDP and the number of observations available for inference. Further research should investigate the reliability of our approach across additional RL algorithms and data sets, as this is essential for addressing real applications challenges. This common effort will enable the development of new research in diverse fields using linked data.

Since the FDP only focuses on falsely linked records, it cannot capture linkage bias: situations where registration errors correlate with certain variables, making it more difficult to recover links involving specific values, thereby impacting inference. While not the focus of our work, synthetic data could also be used in a complementary way to detail missed links and assess linkage bias\cite{linkage_bias_datasynthesis}, extending beyond its role in our FDP estimation procedure. Most likely, methods supporting the development and evaluation of data linkage can benefit from ongoing advances in synthetic data generation and from broader research on model interpretability, which offers tools to better understand statistical mechanisms at stake in RL.

\section{Author contributions statement}
KR implemented the estimation method conceptualised by MW. KR conducted the data applications and the analysis. She prepared the first manuscript draft, reviewed and edited by MH and MW. All authors read and approved this manuscript.

\section{Acknowledgments}
The authors thank the Dutch perinatal health care providers for the Netherlands Perinatal Registry (PRN 11.44), the National Archive of Computerized Data on Aging (NACDA) for their survey (Grant No. U01-AG007198), and the SURF Research Cloud for the computing resources. We also thank the anonymous reviewers for their valuable suggestions.

\section{Supporting information}
The experiments in \texttt{R} and \texttt{Python} and the data sets are available on \href{https://github.com/robachowyk/FDPinRL-experiments}{GitHub}.

\section{Financial disclosure}
This research received no specific grant from any funding agency in the public, commercial, or not-for-profit sectors.

\section{Conflict of interest}
None declared.


\bibliographystyle{unsrt}
\bibliography{reference}

\end{document}